\documentclass[12pt]{article}

\RequirePackage[OT1]{fontenc}
\usepackage{amsthm,amsmath,natbib}
\RequirePackage[colorlinks,citecolor=blue,urlcolor=blue]{hyperref}
\usepackage{subcaption}
\def\frac#1#2{{\textstyle{#1\over#2}}}

\DeclareSymbolFont{AMSb}{U}{msb}{m}{n}
\DeclareMathSymbol{\Natural}{\mathbin}{AMSb}{"4E}
\DeclareMathSymbol{\Integer}{\mathbin}{AMSb}{"5A}
\DeclareMathSymbol{\Real}{\mathbin}{AMSb}{"52}
\DeclareMathSymbol{\Rational}{\mathbin}{AMSb}{"51}
\DeclareMathSymbol{\Imaginary}{\mathbin}{AMSb}{"49}
\DeclareMathSymbol{\Complex}{\mathbin}{AMSb}{"43} 
\DeclareMathSymbol{\Disk}{\mathbin}{AMSb}{"44} 
\def\bi{\begin{itemize}}
\def\ei{\end{itemize}}
\def\bd{\begin{description}}
\def\ed{\end{description}}
\def\ben{\begin{enumerate}}
\def\een{\end{enumerate}}
\def\bv{\begin{verbatim}}
\def\ev{\end{verbatim}}

\def\ith{{$i{\rm th}$ }}
\def\jth{{$j{\rm th}$ }}

\def\calD{{\mathcal D}}

\def\calN{{\mathcal N}}

\def\calP{{\mathcal{P}}}

\def\calS{{\mathcal{S}}}

\def\hat#1{{\widehat{#1}}}

\newcommand{\bs}{\boldsymbol}
\def\pr{{\rm Pr}}

\def\E{{\rm E}}

\def\var{{\rm var}}

\def\2to{{\ {\buildrel 2\over \longrightarrow}\ }}

\def\dotsim{{\ {\buildrel \cdot\over \sim}\ }}
\def\I1ton{{$I_1,\ldots,I_n$}}
\def\X1ton{{$X_1,\ldots,X_n$}}
\def\Y1ton{{$Y_1,\ldots,Y_n$}}
\def\Z1ton{{$Z_1,\ldots,Z_n$}}
\def\R1ton{{$R_1,\ldots,R_n$}}
\def\e1ton{{$e_1,\ldots,e_n$}}
\def\t1ton{{$t_1,\ldots,t_n$}}
\def\x1ton{{$x_1,\ldots,x_n$}}
\def\y1ton{{$y_1,\ldots,y_n$}}
\def\z1ton{{$z_1,\ldots,z_n$}}
\def\calP{{\mathcal{P}}}
\def\calS{{\mathcal{S}}}

\usepackage[english]{babel}
\usepackage{amsfonts}
\usepackage{amssymb}
\usepackage{graphicx}
\usepackage{enumerate}

\begin{document}
\thispagestyle{empty}
\baselineskip=28pt
\vskip 5mm
\begin{center} {\Large{\bf Non-Stationary Dependence Structures for Spatial Extremes}}
\end{center}

\baselineskip=12pt
\vskip 5mm

\begin{center}
\large
Rapha\"el Huser\footnotemark[1], Marc G. Genton{\footnotemark[1]}
\end{center}

\footnotetext[1]{
\baselineskip=10pt CEMSE Division, King Abdullah University of Science and Technology, Thuwal 23955-6900, Saudi Arabia. E-mails: raphael.huser@kaust.edu.sa, marc.genton@kaust.edu.sa}

\baselineskip=17pt
\vskip 4mm
\centerline{\today}
\vskip 6mm

%%%%%%%%%%%%%%%%%%%%%%%%%%%%%%%%%%%%%%%%%%%%%%%%%%%%%%%%%%%%%%%%%%%%%%%%
\begin{center}
{\large{\bf Abstract}}
\end{center}
Max-stable processes are natural models for spatial extremes because they provide suitable asymptotic approximations to the distribution of maxima of random fields. In the recent past, several parametric families of stationary max-stable models have been developed, and fitted to various types of data. However, a recurrent problem is the modeling of non-stationarity. In this paper, we develop non-stationary max-stable dependence structures in which covariates can be easily incorporated. Inference is performed using pairwise likelihoods, and its performance is assessed by an extensive simulation study based on a non-stationary locally isotropic extremal $t$ model. Evidence that unknown parameters are well estimated is provided, and estimation of spatial return level curves is discussed. The methodology is demonstrated with temperature maxima recorded over a complex topography. Models are shown to satisfactorily capture extremal dependence.

\baselineskip=16pt


\pagenumbering{arabic}
\baselineskip=24pt

%%%%%%%%%%%%%%%%%%%%%%%%%%%%%%%%%
%%%%%%%%%%%%%%%%%%%%%%%%%%%%%%%%%
%%%%%%%%%%%%%%%%%%%%%%%%%%%%%%%%%

\section{Introduction} \label{sec.in}
\begin{figure}[t!]
\centering
\includegraphics[width=5.8in]{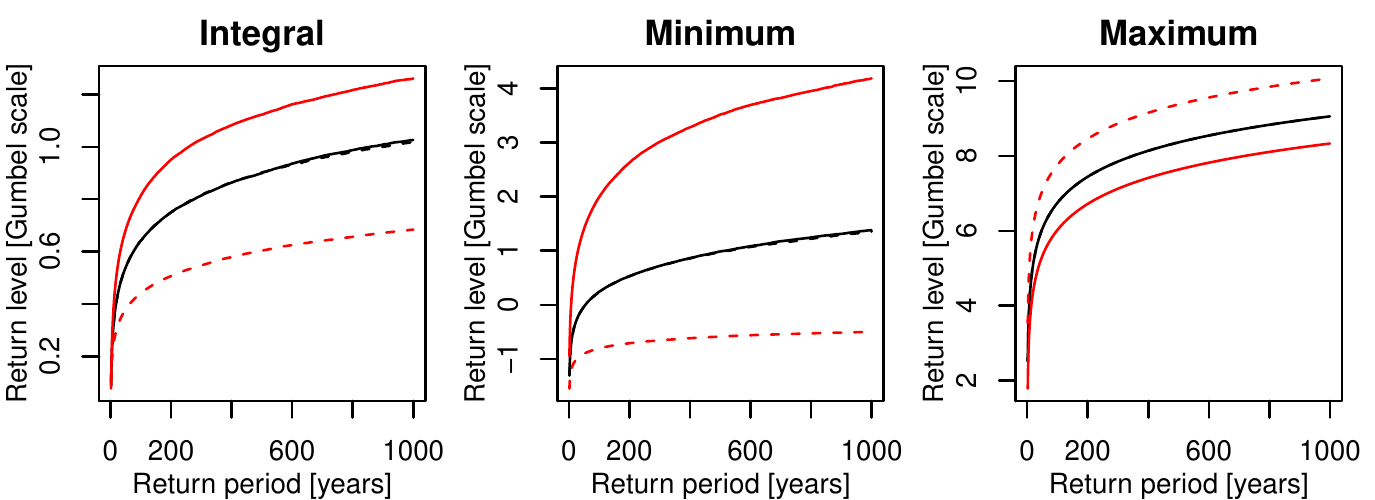}
\caption{True return level curves for the spatial functionals ${\rm INT}_j$ (left), ${\rm MIN}_j$ (middle) and ${\rm MAX}_j$ (right), $j=1,2$, for domains $\calS_1=[0,0.2]\times[0,1]$ (solid) and $\calS_2=[0.8,1]\times[0,1]$ (dashed), based on the extremal $t$ model. Black curves correspond to the stationary case, and red curves to the strongly non-stationary case; more details are given in Section~\ref{sec.sim}.}\label{Fig:ReturnLevels}
\end{figure}

Max-stable processes have drawn attention in the recent past, by providing an asymptotically justified framework for modeling spatial extremes, and allowing extrapolation beyond observed data \citep[see, e.g.,][]{Davison.etal:2012}. Although max-stable processes cannot be characterized by a parametric family, the canonical approach is to fit flexible parametric max-stable models. However, in practice, strong constraints are usually imposed: the max-stable models considered up to now are usually stationary (i.e., shift-invariant) and isotropic (i.e., rotation-invariant). Neglecting non-stationarity at extreme levels may not only provide a poor description of the data, but more importantly, it may also have dramatic consequences on the estimation of return levels (i.e., extrapolation to high quantiles) for spatial quantities, as illustrated by Figure~\ref{Fig:ReturnLevels}. 
%% Figure 1
%\begin{figure}[t!]
%\centering
%\includegraphics[width=5.8in]{ReturnLevels.pdf}
%\caption{True return level curves for the spatial functionals ${\rm INT}_j$ (left), ${\rm MIN}_j$ (middle) and ${\rm MAX}_j$ (right), $j=1,2$, for domains $\calS_1=[0,0.2]\times[0,1]$ (solid) and $\calS_2=[0.8,1]\times[0,1]$ (dashed), based on the extremal $t$ model. Black curves correspond to the stationary case, and red curves to the strongly non-stationary case; more details are given in Section~\ref{sec.sim}.}\label{Fig:ReturnLevels}
%\end{figure}
While it is relatively straightforward to construct non-stationary models for marginal distributions, e.g., by letting the underlying parameters depend on covariates or splines \citep{ChavezDemoulin.Davison:2005,Cooley.etal:2007,Northrop.Jonathan:2011,Davison.Gholamrezaee:2012}, it is more difficult to model non-stationarity in the dependence structure. Furthermore, even if a suitable family of non-stationary models can be identified, performing inference may be awkward if the dataset is not spatially rich enough. Since rare events are scarce by nature, it is even more tricky to detect non-stationary patterns at extreme levels, and there have been very few attempts to tackle this important issue so far. A related problem is the incorporation of substantive knowledge, e.g., from physical processes, into max-stable processes. In particular, information might be gained by including meaningful covariates in the dependence structure.

In an analysis of extreme snow depths, \citet{Blanchet.Davison:2011} proposed splitting the region of study into distinct homogeneous climatic zones to which stationary models were fitted separately, and where anisotropy was dealt with simple geometric deformations of the space. Although their approach simplifies the problem at first sight, it yields a physically unrealistic description of extreme events at the boundary between zones, while the number of parameters also increases dramatically. Another solution advocated by \citet{Cooley.etal:2007} is to map the original latitude-longitude space to an alternative ``climate space'' in which stationarity may be a reasonable assumption, but this might lead to unrealistic realizations and conclusions in the original space. Alternatively, \citet{Smith.Stephenson:2009} and \citet{Reich.Shaby:2012a} proposed Bayesian non-stationary max-stable models. The latter are, however, intrinsically linked to the \citet{Smith:1990b} model, which is built from very smooth storm profiles and therefore lacks flexibility (though the Reich--Shaby model cures this somewhat by having an additional parameter controlling the amount of noise). Furthermore, Bayesian max-stable models are difficult to fit \citep{Ribatet.etal:2012}, although \citet{Thibaud.etal:2015} recently showed how this may be performed in relatively moderate dimensions. In the bivariate case, \citet{deCarvalho.Davison:2014} proposed a non-parametric approach linking different spectral densities through exponential tilting. \citet{Castro.etal:2015} extended this to covariate-dependent spectral densities; see also \citet{deCarvalho:2015}. However, these methods are computationally intensive and difficult to apply in large dimensions.

In the classical geostatistics literature, several non-stationary models have been suggested. \citet{Paciorek.Schervish:2006} proposed a large family of non-stationary correlation functions based on Gaussian kernel convolutions, which can be constructed from known stationary isotropic models. \citet{Nychka.etal:2002} built flexible non-stationary covariance functions using multi-resolution wavelets. \citet{Fuentes:2001} and \citet{Reich.etal:2011b} created non-stationary models by mixing stationary covariance functions and letting the weights depend on covariates. \citet{Jun.Stein:2007,Jun.Stein:2008}, \citet{Castruccio.Stein:2013} and \citet{Castruccio.Genton:2016} advocated a spectral approach that provides flexible non-stationary covariance models on the sphere. Alternatively, \citet{Sampson.Guttorp:1992}, \citet{Perrin.Monestiez:1999}, \citet{Schmidt.OHagan:2003} and \citet{Anderes.Stein:2008} created non-stationary processes by smooth deformations of isotropic random fields. \citet{Bornn.etal:2012} proposed modeling non-stationarity through dimension expansion.  \citet{Lindgren.etal:2011} developed non-stationary models for Gaussian random fields and Gaussian Markov random fields based on stochastic partial differential equations (SPDEs). 

The present paper aims at merging ideas from extreme-value theory and classical geostatistics by proposing simple parametric models able to capture non-stationary patterns in spatial extremes through covariates. To this end, a flexible approach based on max-stable processes and \citeauthor{Paciorek.Schervish:2006}'s correlation model is advocated. Loosely speaking, the new models proposed here are formed by a first layer justified for extremes, within which non-stationarity is handled with locally elliptical kernels, and by a second layer, where these kernels are further described using covariates. As will be explained below, these models can also be seen locally as smoothly deformed isotropic max-stable random fields. Use of mixtures is advocated to capture different smoothness behaviors in distinct subregions.

The full likelihood for max-stable processes is intractable when the number of sites exceeds $D=13$ \citep[see][]{Castruccio.etal:2016}, and for some models, the joint density can only be computed for dimension $D=2$. This explains why pairwise likelihoods \citep{Lindsay:1988,Varin.etal:2011} have become the standard tool for inference in this context \citep{Padoan.etal:2010,Thibaud.etal:2013,Huser.Davison:2014a}, although more efficient approaches based on the point process characterization of extremes have recently been proposed \citep{Wadsworth.Tawn:2014,Engelke.etal:2015,Thibaud.Opitz:2015,Thibaud.etal:2015}.

In Section~\ref{sec.max}, max-stable processes are introduced and some properties and limitations of the Smith--Stephenson model are discussed. In Section~\ref{sec.mod}, we propose new non-stationary max-stable models that are more flexible than the Smith--Stephenson model. In Section~\ref{sec.inf}, we discuss inference based on pairwise likelihoods and in Section~\ref{sec.sim}, we conduct a simulation study to investigate the ability of the estimators to capture nonstationarity in the dependence structure. We also investigate the effect of ignoring non-stationarity on the estimation of spatial return levels. In Section~\ref{sec.ill}, we illustrate the methods on temperature annual maxima recorded in Colorado during $1895$-$1997$, and we conclude with a discussion in Section~\ref{sec.dis}.

%%%%%%%%%%%%%%%%%%%%%%%%%%%%%%%%%
%%%%%%%%%%%%%%%%%%%%%%%%%%%%%%%%%
%%%%%%%%%%%%%%%%%%%%%%%%%%%%%%%%%

\section{Max-stable processes} \label{sec.max}

%%%%%%%%%%%%%%%%%%%%%%%%%%%%%%%%%
%%%%%%%%%%%%%%%%%%%%%%%%%%%%%%%%%

\subsection{Theoretical foundation}

%Suppose that $X_1,X_2,\ldots,$ are independent and identically distributed (i.i.d.) random variables with distribution $F(x)$. If there exist normalizing sequences $a_n>0$ and $b_n$ such that the renormalized maximum $a_n^{-1}\{\max(X_1,\ldots,X_n)-b_n\}$ converges to a non-degenerate variable $Z\sim G(z)$, as $n\to\infty$, then the limiting distribution $G(z)$ must be generalized extreme-value (GEV), or equivalently \emph{max-stable}, i.e., $G(z)=\exp\left[-\left\{1+\xi(z-\mu)/\sigma\right\}_+^{-1/\xi}\right]$, where $a_+=\max(0,a)$, and $\mu,\sigma>0,\xi$ are the location, scale and shape parameters, respectively \citep{Coles:2001}. This result provides strong support for fitting the GEV to block maxima, though the asymptotic approximation applied to finite $n$ entails several practical difficulties; see, e.g., the review by \citet{Davison.Huser:2015}. Furthermore, by standardizing the individual variables as $Y_i=1/\{1-F(X_i)\}$ $(i=1,2,\ldots)$, we can show that $n^{-1}\max(Y_1,\ldots,Y_n)$ converges to a GEV distribution with $\mu=\sigma=\xi=1$ (unit Fr\'echet distribution).

Suppose that $X_1(\bs{s}),X_2(\bs{s}),\ldots$, are independent and identically distributed random processes with continuous sample paths on $\calS\subset\Real^d$, and that there exist sequences of functions $a_n(\bs{s})>0$ and $b_n(\bs{s})$ such that the renormalized process of pointwise maxima $a_n(\bs{s})^{-1}[\max\{X_1(\bs{s}),\ldots,X_n(\bs{s})\}-b_n(\bs{s})]$ converges weakly to a process $Z(\bs{s})$ with non-degenerate margins, as $n\to\infty$. Then, $Z(\bs{s})$ must be max-stable, i.e., for any positive integer $k$, the finite-dimensional distributions of $Z(\bs{s})$ and $\max\{Z_1(\bs{s}),\ldots,Z_k(\bs{s})\}$, where $Z_1(\bs{s}),\ldots,Z_k(\bs{s})$ denote independent replicates of $Z(\bs{s})$, differ only through location and scale coefficients. In particular, margins follow the generalized extreme-value distribution $G(z)=\exp\left(-\left[1+\xi(\bs{s})\{z-\mu(\bs{s})\}/\sigma(\bs{s})\right]_+^{-1/\xi(\bs{s})}\right)$, with spatially-varying location, scale and shape parameters $\mu(\bs{s}),\sigma(\bs{s})>0,\xi(\bs{s})$, respectively. Furthermore, defining standardized processes as $Y_i(\bs{s})=1/[1-F_{\bs{s}}\{X_i(\bs{s})\}]$ $(i=1,2,\ldots)$ with $F_{\bs{s}}(x)$ the marginal distribution of $X(\bs{s})$ at location $\bs{s}$, the limiting distribution of $n^{-1}\max\{Y_1(\bs{s}),\ldots,Y_n(\bs{s})\}$ is max-stable with unit Fr\'echet margins (i.e., GEV with parameters $\mu(\bs{s})=\sigma(\bs{s})=\xi(\bs{s})=1$). Such a limiting process is called simple max-stable. Standardization allows the treatment of the margins to be separated from the dependence structure.

Simple max-stable processes have been characterized by \citet{deHaan:1984}; see also \citet{Schlather:2002} and \citet[][\S9.4]{deHaan.Ferreira:2006}. Given points $\{P_i;i=1,2,\ldots\}$ of a Poisson process with intensity $p^{-2}$ $(p>0)$ and independent replicates $\{W_i(\bs{s});i=1,2,\ldots\}$ of a positive process $W(\bs{s})$ $(\bs{s}\in\calS\subset\Real^d)$ with unit mean, the process created as
\begin{equation}\label{characMS}
Z(\bs{s})=\sup_{i=1,2,\ldots} P_i W_i(\bs{s})
\end{equation}
is a simple max-stable process. Conversely, under mild conditions, each continuous simple max-stable process can be decomposed as in  (\ref{characMS}). Furthermore, for any set of $D$ spatial locations $\calD=\{\bs{s}_1,\ldots,\bs{s}_D\}\subset\calS$, one has
\begin{equation}\label{jointD}
\pr\{Z(\bs{s}_1)\leq z_1,\ldots,Z(\bs{s}_D)\leq z_D\}=\exp\left\{-V_{\calD}\left(z_1,\ldots,z_D\right)\right\},
\end{equation}
where the so-called exponent measure is $V_{\calD}\left(z_1,\ldots,z_D\right)=\E\left[\max\left\{{W(\bs{s}_1)/ z_1},\ldots,{W(\bs{s}_D)/ z_D}\right\}\right]$. The exponent measure has a closed-form formula for specific choices of $W(\bs{s})$; see, e.g., \citet{Schlather:2002}, \citet{Nikoloulopoulos.etal:2009}, \citet{Genton.etal:2011}, \citet{Huser.Davison:2013a}, and \citet{Opitz:2013a}. A useful related quantity is the so-called extremal coefficient $\theta(\bs{s}_1,\bs{s}_2)=V_\calD(1,1)\in[1,2]$, $\calD=\{\bs{s}_1,\bs{s}_2\}$, giving a measure of dependence between variables $Z(\bs{s}_1)$ and $Z(\bs{s}_2)$, or equivalently, extremal dependence between variables $Y(\bs{s}_1)$ and $Y(\bs{s}_2)$: $\theta(\bs{s}_1,\bs{s}_2)=1$ corresponds to perfect dependence and $\theta(\bs{s}_1,\bs{s}_2)=2$ to independence.

For more details about univariate and multivariate extremes, see \citet{Beirlant.etal:2004} and \citet{Davison.Huser:2015}, and for an account of spatial extremes, see the review papers by \citet{Davison.etal:2012}, \citet{Cooley.etal:2012a} and \citet{Davison.etal:2013}. See also the book by \citet{deHaan.Ferreira:2006}, which explains the technicalities in depth.

%%%%%%%%%%%%%%%%%%%%%%%%%%%%%%%%%
%%%%%%%%%%%%%%%%%%%%%%%%%%%%%%%%%

\subsection{The celebrated Smith model and its non-stationary extension}\label{smi.ext}

\begin{figure}[t!]
\centering
\includegraphics[width=0.65\linewidth]{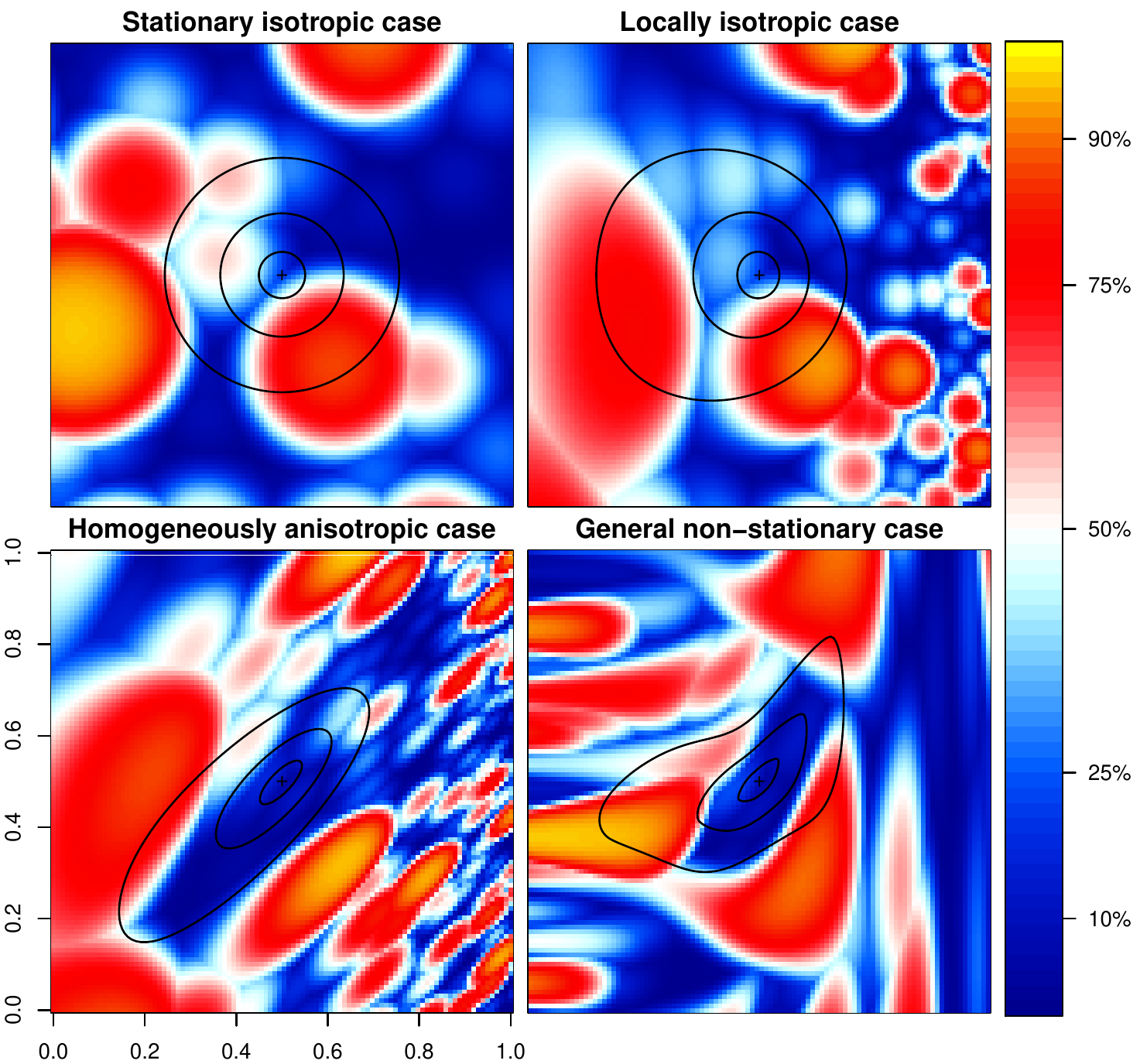}
\caption{Simulations of Model (\ref{NonStatSmith}) for locations $\bs{s}=(s_x,s_y)\in[0,1]^2$. \emph{Top left}: stationary isotropic case with $\bs{\Omega}_{\bs{s}}=0.1^2\bs{I}_2$.  \emph{Top right}: non-stationary locally isotropic case with $\bs{\Omega}_{\bs{s}}=0.4^22^{-8|s_x|}\bs{I}_2$. \emph{Bottom left}: non-stationary homogeneously anisotropic case with $\bs{\Omega}_{\bs{s}}=0.4^22^{-8|s_x|}\bs{R}$, $\bs{R}\in\Real^{2\times2}$ being a correlation matrix with correlation $0.8$. \emph{Bottom right}: general non-stationary case with $(\bs{\Omega}_{\bs{s}})_{11}=0.4^22^{-8|s_x|}$, $(\bs{\Omega}_{\bs{s}})_{22}=0.4^22^{-8|1-s_x|}$ and $(\bs{\Omega}_{\bs{s}})_{12}=(\bs{\Omega}_{\bs{s}})_{21}=\{(\bs{\Omega}_{\bs{s}})_{11}(\bs{\Omega}_{\bs{s}})_{22}\}^{1/2}\{e^{h(\bs{s})}-1\}/\{e^{h(\bs{s})}+1\}$, $h(\bs{s})=2\log(3)e^{-30(s_x-0.5)^2}$. Realizations are based on the same random seed. The contours correspond to $\theta(\bs{s}_1,\bs{s}_2)=1.2,1.5,1.8$ (narrow to wide), where $\bs{s}_1$ is the center location (cross). The color scale indicates quantile probabilities.
}\label{Fig:NonStatSmith}
\end{figure}

The first stationary max-stable model proposed in the literature is the \citet{Smith:1990b} model, which assumes in (\ref{characMS}) that $W_i(\bs{s})=\phi_d(\bs{s}-\bs{U}_i;\bs{\Omega})$, where the $\bs{U}_i$s are the points of a unit rate Poisson process on $\calS=\Real^d$ and $\phi_d(\cdot;\bs{\Omega})$ denotes the $d$-dimensional Gaussian density function with covariance matrix $\bs{\Omega}$. Although finite-dimensional distributions are known in arbitrary dimensions \citep{Genton.etal:2011}, they are always degenerate for $D>d+1$, which raises the question of the suitability of the Smith model in practice. The non-stationary extension proposed by \citet{Smith.Stephenson:2009} considers spatially varying covariance matrices $\bs{\Omega}_{\bs{s}}$, capturing the small-scale dependence structure around location $\bs{s}\in\calS$. The generalized storm profiles are of the form
\begin{equation}\label{NonStatSmith}
W_i(\bs{s})=\phi_d(\bs{s}-\bs{U}_i;\bs{\Omega}_{\bs{U}_i}).
\end{equation}
This model has the appealing property of being locally elliptic (a feature that we will retain for the more general model proposed in Section~\ref{sec.mod}), in the sense that infinitesimal contours of the extremal coefficient form ellipses, see Figure~\ref{Fig:NonStatSmith}. Several special cases may be of interest in practice: if contours are locally circular with $\bs{\Omega}_{\bs{s}}=\omega^2({\bs{s}})\bs{I}_d$, where $\omega({\bs{s}})>0$ and $\bs{I}_d$ is the $d$-by-$d$ identity matrix, the model is locally isotropic (top right panel of Figure~\ref{Fig:NonStatSmith}), and when $\omega({\bs{s}})=\omega>0$ for all $\bs{s}\in\calS$, (\ref{NonStatSmith}) reduces to the stationary isotropic case, i.e., the classical Smith model (top left panel of Figure~\ref{Fig:NonStatSmith}). When $\bs{\Omega}_{\bs{s}}=\omega^2({\bs{s}})\bs{R}$ for some fixed $d$-by-$d$ correlation matrix $\bs{R}$, the model is not isotropic, but the anisotropy is homogeneous over space; see the bottom left panel of Figure~\ref{Fig:NonStatSmith}. If $\omega({\bs{s}})=\omega>0$ for all $\bs{s}\in\calS$, it reduces to the stationary anisotropic case, illustrated by \citet{Blanchet.Davison:2011}. \citet{Smith.Stephenson:2009} provide bivariate margins in the homogeneously anisotropic case only; in the Supplementary Material, calculations are performed in full generality for $D=2$.

%% Figure 2
%\begin{figure}[t!]
%\centering
%\includegraphics[width=0.65\linewidth]{SimSmith2.pdf}
%\caption{Simulations of Model (\ref{NonStatSmith}) for locations $\bs{s}=(s_x,s_y)\in[0,1]^2$. \emph{Top left}: stationary isotropic case with $\bs{\Omega}_{\bs{s}}=0.1^2\bs{I}_2$.  \emph{Top right}: non-stationary locally isotropic case with $\bs{\Omega}_{\bs{s}}=0.4^22^{-8|s_x|}\bs{I}_2$. \emph{Bottom left}: non-stationary homogeneously anisotropic case with $\bs{\Omega}_{\bs{s}}=0.4^22^{-8|s_x|}\bs{R}$, $\bs{R}\in\Real^{2\times2}$ being a correlation matrix with correlation $0.8$. \emph{Bottom right}: general non-stationary case with $(\bs{\Omega}_{\bs{s}})_{11}=0.4^22^{-8|s_x|}$, $(\bs{\Omega}_{\bs{s}})_{22}=0.4^22^{-8|1-s_x|}$ and $(\bs{\Omega}_{\bs{s}})_{12}=(\bs{\Omega}_{\bs{s}})_{21}=\{(\bs{\Omega}_{\bs{s}})_{11}(\bs{\Omega}_{\bs{s}})_{22}\}^{1/2}\{e^{h(\bs{s})}-1\}/\{e^{h(\bs{s})}+1\}$, $h(\bs{s})=2\log(3)e^{-30(s_x-0.5)^2}$. Realizations are based on the same random seed. The contours correspond to $\theta(\bs{s}_1,\bs{s}_2)=1.2,1.5,1.8$ (narrow to wide), where $\bs{s}_1$ is the center location (cross). The color scale indicates quantile probabilities.
%}\label{Fig:NonStatSmith}
%\end{figure}

The extremal coefficient of the stationary Smith model satisfies $\theta(\bs{s}_1,\bs{s}_2)\equiv \theta(\|\bs{h}\|)\to2$, as $\|\bs{h}\|=\|\bs{s}_1-\bs{s}_2\|\to\infty$, which implies that complete independence can be captured at infinity. In $\Integer$, this is equivalent to the process being \emph{mixing} \citep{Kabluchko.Schlather:2010}. In the Supplementary Material, we show that this property is also fulfilled by the Smith--Stephenson model with $\Omega_{\bs{s}}=\omega({\bs{s}})^2 \bs{I}_d$ (locally isotropic case) provided $\omega(\bs{s}) = o(\|\bs{s}\|)$; by a simple extension, this is also true when $\bs{\Omega}_{\bs{s}}=\omega({\bs{s}})^2 \bs{R}$ with $\bs{R}$ a correlation matrix (homogeneously anisotropic case). This result makes sense because if one has $\omega({\bs{s}})=O(\|\bs{s}\|)$, the extent of a storm centered at $\bs{s}$ increases at the same rate as the distance separating $\bs{s}$ from any fixed other point $\bs{s}_0$, such that the storm contributes to the supremum (\ref{characMS}) at location $\bs{s}_0$ with positive probability, no matter how far it is from $\bs{s}_0$. 

Although the Smith--Stephenson model is easily interpretable, it has several limitations. First, finite-dimensional distributions are known for $D=2$ only. Second, pairwise densities involve the cumulative distribution and density of quadratic forms of normal variables, the computation of which may be intensive (see the Supplementary Material). Finally, as illustrated in Figure~\ref{Fig:NonStatSmith}, this process is very smooth. Realizations are infinitely differentiable in neighborhoods of all points that do not lie on the border between distinct storms, and this appears too strong an assumption in most environmental applications. In fact, the storm profiles are almost deterministic; randomness is solely created by the storm locations $\bs{U}_i$ and storm intensities $P_i$  in (\ref{characMS}). More flexible non-stationary max-stable models with stochastic storm profiles, generalizing \eqref{NonStatSmith}, are proposed in Section~\ref{sec.mod}.

%%%%%%%%%%%%%%%%%%%%%%%%%%%%%%%%%
%%%%%%%%%%%%%%%%%%%%%%%%%%%%%%%%%
%%%%%%%%%%%%%%%%%%%%%%%%%%%%%%%%%

\section{Flexible non-stationary dependence structures} \label{sec.mod}

%%%%%%%%%%%%%%%%%%%%%%%%%%%%%%%%%
%%%%%%%%%%%%%%%%%%%%%%%%%%%%%%%%%
\subsection{The non-stationary extremal $t$ model}\label{sec.ext}
The extremal $t$ model \citep{Nikoloulopoulos.etal:2009,Opitz:2013a} is defined by taking
\begin{equation}\label{ExtremalT}
W(\bs{s})=c_{\rm df}\max\{0,\varepsilon(\bs{s})\}^{\rm df}, \qquad c_{\rm df}=2^{1-{\rm df/2}}\pi^{1/2}\left[\Gamma\left\{({{\rm df}+1)/ 2}\right\}\right]^{-1},
\end{equation}
in (\ref{characMS}), where ${\rm df}>0$, $\varepsilon(\bs{s})$ is a Gaussian process with zero mean, unit variance and correlation function $\rho(\bs{s}_1,\bs{s}_2)$, and $\Gamma(\cdot)$ is the gamma function. The extremal $t$ model does not capture independence unless ${\rm df}\to\infty$ \citep{Davison.etal:2012}, but this issue may be resolved by incorporating a random set element \citep{Davison.Gholamrezaee:2012,Huser.Davison:2014a}, though the inference is more tricky. The model \eqref{ExtremalT} has several interesting sub-models, the \emph{stationary} versions of which have been applied extensively. When ${\rm df}=1$, (\ref{ExtremalT}) reduces to the \citet{Schlather:2002} model, which has been fitted in numerous applications \citep{Davison.Gholamrezaee:2012,Davison.etal:2012,Ribatet:2013b,Thibaud.etal:2013}.  The Brown--Resnick process \citep{Brown.Resnick:1977,Kabluchko.etal:2009} arises as a limiting case of (\ref{ExtremalT}) as ${\rm df}\to\infty$ \citep{Davison.etal:2012}; its storm profiles may be expressed as $W(\bs{s})=\exp\{\varepsilon(\bs{s})-\gamma(\bs{s})\}$, where $\varepsilon(\bs{s})$ is a Gaussian random field with semi-variogram $\gamma(\bs{h})$ such that $\varepsilon(\bs{0})=0$ almost surely. The Brown--Resnick process extends the Smith model \citep{Huser.Davison:2013a}, and it can also be viewed as the generalization of the \citet{Husler.Reiss:1989} multivariate extreme-value distribution to the spatial framework. In practice, Brown--Resnick processes have proven to be quite flexible compared to the Smith and Schlather alternatives \citep{Davison.etal:2012,Jeon.Smith:2012}. Model (\ref{ExtremalT}) not only generalizes all aforementioned stationary max-stable models, but it is also the max-attractor for the broad class of all suitably rescaled elliptical processes \citep{Opitz:2013a}, which provides strong support for its use in practice; as an illustration of its practical performance, see \citet{Thibaud.Opitz:2015}. The bivariate exponent measure for (\ref{ExtremalT}) may be expressed as
\begin{equation}
V_{\calD}\left(z_1,z_2\right)={1\over z_1}T_{{\rm df}+1}\left[{\left({z_2/ z_1}\right)^{1/{\rm df}}-\rho(\bs{s}_1,\bs{s}_2)\over\left({\rm df}+1\right)^{-1/2}\left\{1-\rho(\bs{s}_1,\bs{s}_2)^2\right\}^{1/2}}\right] + {1\over z_2}T_{{\rm df}+1}\left[{\left({z_1/ z_2}\right)^{1/{\rm df}}-\rho(\bs{s}_1,\bs{s}_2)\over\left({\rm df}+1\right)^{-1/2}\left\{1-\rho(\bs{s}_1,\bs{s}_2)^2\right\}^{1/2}}\right],\label{ExponentMeasureExtremalT}
\end{equation}
where $T_{\rm df}(\cdot)$ is the Student $t$ cumulative distribution function with ${\rm df}$ degrees of freedom. Explicit expressions in dimension $D$ are also available \citep[see][]{Thibaud.Opitz:2015}. 

\begin{figure}[t!]
\centering
\includegraphics[width=0.85\linewidth]{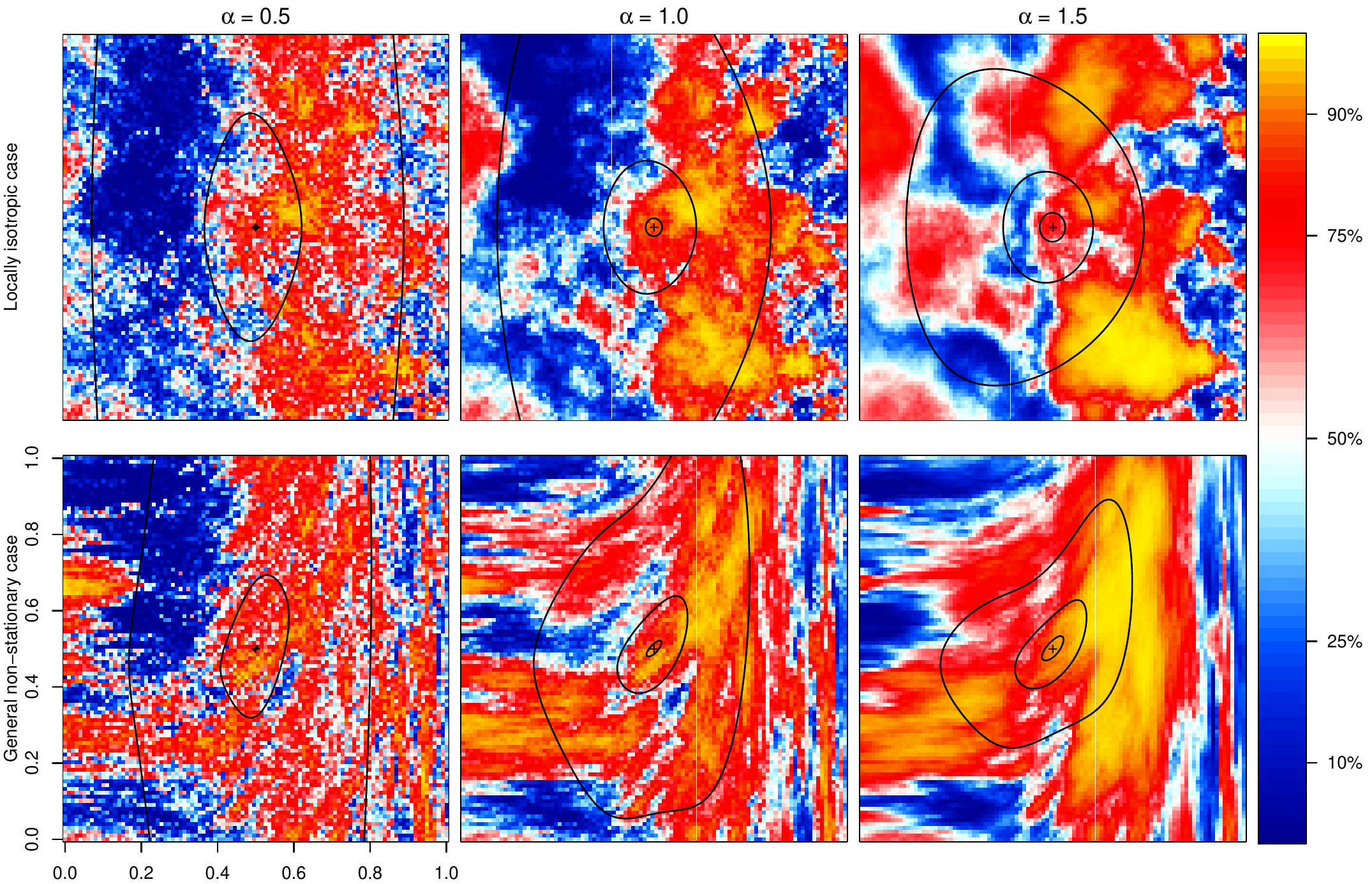}
\caption{Simulation of the extremal $t$ model (\ref{ExtremalT}), with ${\rm df=5}$ and non-stationary correlation function (\ref{PaciorekCorrelation}), combined with (\ref{PoweredExponential}), in $[0,1]^2$. Columns correspond to different smoothness scenarios, with $\alpha=0.5,1,1.5$ (left to right). Locally isotropic (top row) and general non-stationary (bottom row) cases are displayed. The underlying spatially-varying matrices are $\bs{\Omega}_{\bs{s}}=(2\,{\rm df})^{2/\alpha}\times\bs{\Omega}_{\bs{s}}^{\rm BR}$, where $\bs{\Omega}_{\bs{s}}^{\rm BR}=0.4^22^{-8|s_x|}\bs{I}_2$ (top row) or $(\bs{\Omega}_{\bs{s}}^{\rm BR})_{11}=0.4^22^{-8|s_x|}$, $(\bs{\Omega}_{\bs{s}}^{\rm BR})_{22}=0.4^22^{-8|1-s_x|}$ and $(\bs{\Omega}_{\bs{s}}^{\rm BR})_{12}=(\bs{\Omega}_{\bs{s}}^{\rm BR})_{21}=\{(\bs{\Omega}_{\bs{s}}^{\rm BR})_{11}(\bs{\Omega}_{\bs{s}}^{\rm BR})_{22}\}^{1/2}\{e^{h(\bs{s})}-1\}/\{e^{h(\bs{s})}+1\}$, where $h(\bs{s})=2\log(3)e^{-30(s_x-0.5)^2}$ (bottom row). Realizations are from the same random seed. Contour curves correspond to $\theta(\bs{s}_1,\bs{s}_2)=1.2,1.5,1.8$ (narrow to wide), where $\bs{s}_1$ is the center location (cross). The color scale indicates quantile probabilities.
}\label{Fig:NonStatExtremalT}
\end{figure}

Our approach to modeling non-stationarity in spatial extremes consists of combining the extremal $t$ model (\ref{ExtremalT}) with non-stationary correlation functions $\rho(\bs{s}_1,\bs{s}_2)$ proposed in the classical spatial statistics literature. As mentioned above, there exist several ways to construct non-stationary correlation functions, spanning from space deformations to SPDEs, and including wavelets, spectral methods, mixtures of stationary correlations or kernel convolutions. Hence, our methodology to tackle non-stationarity in extremes is very general and can potentially yield a large variety of models, having their own advantages and drawbacks. There are (at least) three desirable properties that we would like our model to possess: simplicity, local ellipticity, and ease to incorporate covariates. We have found that the kernel convolution approach advocated by \citet{Paciorek.Schervish:2006} is especially satisfactory. These authors have proposed a very general construction of non-stationary correlation functions that are based on known isotropic correlation models. Specifically, let $\bs{\Omega}_{\bs{s}}$ denote a (continuously) spatially varying $d$-by-$d$ covariance matrix, and for any two locations $\bs{s}_1,\bs{s}_2\in\calS$ with separation vector $\bs{h}=\bs{s}_2-\bs{s}_1$, define the quadratic form  $Q_{\bs{s}_1;\bs{s}_2}$ as
\begin{equation}\label{QuadraticForm}
Q_{\bs{s}_1;\bs{s}_2}=\bs{h}^T\left({\bs{\Omega}_{\bs{s}_1}+\bs{\Omega}_{\bs{s}_2}\over 2}\right)^{-1}\bs{h}.
\end{equation}
\citet{Paciorek.Schervish:2006} show that for any isotropic correlation function $R(\|\bs{h}\|)$ valid on $\Real^d$ $(d=1,2,\ldots)$, the function 
\begin{equation}\label{PaciorekCorrelation}
\rho(\bs{s}_1,\bs{s}_2)=|\bs{\Omega}_{\bs{s}_1}|^{1/4}|\bs{\Omega}_{\bs{s}_2}|^{1/4}\bigg|{\bs{\Omega}_{\bs{s}_1}+\bs{\Omega}_{\bs{s}_2}\over 2}\bigg|^{-1/2}R\left({Q_{\bs{s}_1;\bs{s}_2}}^{1/2}\right)
\end{equation}
provides a valid non-stationary correlation function on $\Real^d$ $(d=1,2,\ldots)$. To avoid parametrization redundancy, the function $R(\|\bs{h}\|)$ can be assumed to have unit range. Many isotropic correlation functions have been proposed in the literature (see, e.g., \citealp{Cressie:1993}, \citealp{Stein:1999} or \citealp{Cressie.Wikle:2011}), making \eqref{PaciorekCorrelation} a useful constructive device for non-stationary correlation functions. One popular possibility is the powered exponential family
\begin{equation}\label{PoweredExponential}
R(\|\bs{h}\|)=\exp\left(-\|\bs{h}\|^\alpha\right),
\end{equation}
where $\alpha\in(0,2]$ is a smoothness parameter, and the exponential and squared exponential models correspond to $\alpha=1$ and $\alpha=2$, respectively. This correlation family generates random fields with very rough (with $\alpha\to0$) to analytical sample paths (with $\alpha=2$). Hence, great flexibility can be obtained by combining (\ref{PaciorekCorrelation}) with (\ref{PoweredExponential}). Since the max-stable model in (\ref{ExtremalT}) inherits its sample path differentiability properties from the underlying Gaussian process $\varepsilon(\bs{s})$, the parameter $\alpha$ in (\ref{PoweredExponential}) has a direct relationship with the smoothness of the resulting max-stable process. To illustrate this, typical realizations from the non-stationary extremal $t$ model with ${\rm df}=5$ combined with (\ref{PaciorekCorrelation}) and (\ref{PoweredExponential}) are displayed in Figure~\ref{Fig:NonStatExtremalT}. 
%% Figure 3
%\begin{figure}[t!]
%\centering
%\includegraphics[width=0.85\linewidth]{SimExtremalT2.pdf}
%\caption{Simulation of the extremal $t$ model (\ref{ExtremalT}), with ${\rm df=5}$ and non-stationary correlation function (\ref{PaciorekCorrelation}), combined with (\ref{PoweredExponential}), in $[0,1]^2$. Columns correspond to different smoothness scenarios, with $\alpha=0.5,1,1.5$ (left to right). Locally isotropic (top row) and general non-stationary (bottom row) cases are displayed. The underlying spatially-varying matrices are $\bs{\Omega}_{\bs{s}}=(2\,{\rm df})^{2/\alpha}\times\bs{\Omega}_{\bs{s}}^{\rm BR}$, where $\bs{\Omega}_{\bs{s}}^{\rm BR}=0.4^22^{-8|s_x|}\bs{I}_2$ (top row) or $(\bs{\Omega}_{\bs{s}}^{\rm BR})_{11}=0.4^22^{-8|s_x|}$, $(\bs{\Omega}_{\bs{s}}^{\rm BR})_{22}=0.4^22^{-8|1-s_x|}$ and $(\bs{\Omega}_{\bs{s}}^{\rm BR})_{12}=(\bs{\Omega}_{\bs{s}}^{\rm BR})_{21}=\{(\bs{\Omega}_{\bs{s}}^{\rm BR})_{11}(\bs{\Omega}_{\bs{s}}^{\rm BR})_{22}\}^{1/2}\{e^{h(\bs{s})}-1\}/\{e^{h(\bs{s})}+1\}$, where $h(\bs{s})=2\log(3)e^{-30(s_x-0.5)^2}$ (bottom row). Realizations are from the same random seed. Contour curves correspond to $\theta(\bs{s}_1,\bs{s}_2)=1.2,1.5,1.8$ (narrow to wide), where $\bs{s}_1$ is the center location (cross). The color scale indicates quantile probabilities.
%}\label{Fig:NonStatExtremalT}
%\end{figure}
%It can be noticed from the extremal coefficient contours that rougher random fields are less dependent at short distances as expected.

Like the non-stationary Smith model, the correlation function (\ref{PaciorekCorrelation}) is locally elliptic, and this attractive geometric property is therefore preserved for the resulting non-stationary max-stable random field. This implies that the latter can be seen locally as a smoothly deformed isotropic max-stable process. To see this, fix $\bs{s}_0\in\calS$ and let $\bs{s}_1,\bs{s}_2\in N(\bs{s}_0)\subset\calS$ be two locations within some small neighborhood $N(\bs{s}_0)$ of $\bs{s}_0$. By continuity of the map $\bs{s}\mapsto\bs{\Omega}_{\bs{s}}$, one has that $\bs{\Omega}_{\bs{s}_2}\approx\bs{\Omega}_{\bs{s}_1}\approx\bs{\Omega}_{\bs{s}_0}$ and $Q_{\bs{s}_1;\bs{s}_2}\approx \bs{h}^T \bs{\Omega}_{\bs{s}_0}^{-1} \bs{h}$, where $\bs{h}=\bs{s}_2-\bs{s}_1$ is the lag vector. Then, applying the spatial transformation $\bs{s}\mapsto \bs{s}^\star=\bs{\Omega}_{\bs{s}_0}^{-1/2} (\bs{s}-\bs{s}_0)$ in $N(\bs{s}_0)$, where $\bs{\Omega}_{\bs{s}_0}=\bs{\Omega}_{\bs{s}_0}^{1/2}\bs{\Omega}_{\bs{s}_0}^{T/2}$, one can easily verify that the correlation function on the new coordinate system satisfies $\rho(\bs{s}_1^\star,\bs{s}_2^\star)\approx R(\|\bs{h}^\star\|)$ with $\bs{h}^\star=\bs{s}_2^\star-\bs{s}_1^\star$; it is therefore locally isotropic.

Another appealing feature is that the proposed non-stationary extremal $t$ model defined above using (\ref{PaciorekCorrelation}) and (\ref{PoweredExponential}) with covariance matrices $\bs{\Omega}_{\bs{s}}=(2\,{\rm df})^{2/\alpha}\times\bs{\Omega}_{\bs{s}}^{{\rm BR}}$, converges as ${\rm df}\to\infty$ to the Brown--Resnick process with variogram $2\gamma(\bs{s}_1,\bs{s}_2)=({Q_{\bs{s}_1;\bs{s}_2}^{\rm BR}})^{\alpha/2}$, where $Q_{\bs{s}_1;\bs{s}_2}^{\rm BR}$ is defined in (\ref{QuadraticForm}) using $\bs{\Omega}_{\bs{s}}^{\rm BR}$. In particular, the Smith--Stephenson model (\ref{NonStatSmith}) is recovered when $\alpha=2$. In practice, this implies that it is enough to fit the non-stationary extremal $t$ model, as our approach generalizes (\ref{NonStatSmith}); if ${\rm df}$ is found to be relatively large and $\alpha\approx 2$, then it might also be interesting to consider the Smith--Stephenson model, although it is more complex to fit.

%%%%%%%%%%%%%%%%%%%%%%%%%%%%%%%%%
%%%%%%%%%%%%%%%%%%%%%%%%%%%%%%%%%
\subsection{Covariates}\label{sec.covar}
We now continue our modeling on the plane with ${d=2}$, although our approach could be applied in higher dimensions. In order to retain simplicity in our modeling of non-stationarity, we seek to incorporate meaningful covariates in the extremal dependence structure. To this end, we propose further modeling the covariance matrices $\bs{\Omega}_{\bs{s}}$ $(\bs{s}\in\calS)$ as follows: let
\begin{eqnarray}
\bs{\Omega}_{\bs{s}}&=&\begin{pmatrix}
\omega_x^2(\bs{s}) & \omega_x(\bs{s})\omega_y(\bs{s})\delta(\bs{s})\\
\omega_x(\bs{s})\omega_y(\bs{s})\delta(\bs{s}) & \omega_y^2(\bs{s})
\end{pmatrix},\quad\mbox{with, for example,}\label{MatrixOmega}
\end{eqnarray}
%with, for example,
\begin{equation}\label{CovariatesModeling}
\log\{\omega_x(\bs{s})\}=\bs{X}_{\omega_x}^T(\bs{s})\bs{\beta}_{\omega_x},\;\;\;\log\{\omega_y(\bs{s})\}=\bs{X}_{\omega_y}^T(\bs{s})\bs{\beta}_{\omega_y},\;\;\;{\rm logit}[\{\delta(\bs{s})+1\}/2]=\bs{X}_\delta^T(\bs{s})\bs{\beta}_\delta,
\end{equation}
where $\bs{X}_{\omega_x}(\bs{s}),\bs{X}_{\omega_y}(\bs{s})$ and $\bs{X}_\delta(\bs{s})$ denote vectors of covariates corresponding to location $\bs{s}$, and $\bs{\beta}_{\omega_x},\bs{\beta}_{\omega_y}$ and $\bs{\beta}_\delta$ are the associated vectors of parameters measuring importance of covariates. The link functions in (\ref{CovariatesModeling}) ensure that $\omega_x(\bs{s})>0$, $\omega_y(\bs{s})>0$ and $\delta(\bs{s})\in(-1,1)$, but they could in principle be replaced by other functions that satisfy these conditions. The construction (\ref{MatrixOmega}) guarantees the positive definiteness of $\bs{\Omega}_{\bs{s}}$. 
The local correlation range at station $\bs{s}$ with respect to the $x$ (respectively $y$) axis is measured by the functions $\omega_x(\bs{s})$ (respectively $\omega_y(\bs{s})$), whereas $\delta(\bs{s})$ captures the local anisotropy level: if $\delta(\bs{s})=0$ and $\omega_x(\bs{s})=\omega_y(\bs{s})$, the resulting process is locally isotropic, i.e., infinitesimal contours are circular everywhere, whereas if $\delta(\bs{s})\neq 0$, contours are slanted ellipses; see Figure~\ref{Fig:NonStatExtremalT}.

%%%%%%%%%%%%%%%%%%%%%%%%%%%%%%%%%
%%%%%%%%%%%%%%%%%%%%%%%%%%%%%%%%%
\subsection{Max-stable mixtures}\label{sec.alt}
Although the non-stationary model \eqref{ExtremalT} appears quite flexible, one limitation is that it has a single smoothness parameter for the whole region. This issue may be overcome by using non-stationary Mat\'ern correlation functions \citep{Stein:2005b,Anderes.Stein:2011}, or by using an approach based of mixtures. The latter is outlined below.

The first type of mixture consists of max-mixtures of max-stable models. Let $Z^1(\bs{s})$ and $Z^2(\bs{s})$ be independent max-stable processes with unit Fr\'echet margins defined on the same space $\calS$. Then for any function $0\leq a(\bs{s})\leq 1$, the spatial process defined as $Z(\bs{s})=\max[a(\bs{s})Z^1(\bs{s}),\{1-a(\bs{s})\}Z^2(\bs{s})]$ is a simple max-stable process with exponent measure
\begin{equation}\label{maxmix}
V_\calD(z_1,\ldots,z_D)=a(\bs{s})V_\calD^1(z_1,\ldots,z_D) + \{1-a(\bs{s})\}V_\calD^2(z_1,\ldots,z_D),
\end{equation}
where $V_\calD^1$ and $V_\calD^2$ are the exponent measures of $Z^1(\bs{s})$ and $Z^2(\bs{s})$, respectively. The function $a(\bs{s})$ is a spatially varying proportion, determining which of the processes $Z^1(\bs{s})$ and $Z^2(\bs{s})$ is dominant at location $\bs{s}$. Model (\ref{maxmix}) is stationary if $a(\bs{s})$ is constant over space and $Z^1(\bs{s})$ and $Z^2(\bs{s})$ are stationary, but it can be made non-stationary by allowing $a(\bs{s})$ to depend upon covariates, e.g., ${\rm logit}\{a(\bs{s})\}=\bs{X}_a^T(\bs{s})\bs{\beta}_a$, where $\bs{X}_a(\bs{s})$ is a vector of covariates for location $\bs{s}$ and $\bs{\beta}_a$ is the associated vector of parameters. Different smoothness behaviors may be captured in different spatial regions, provided $Z^1(\bs{s})$ and $Z^2(\bs{s})$ have different degrees of differentiability. More complex non-stationary max-stable models $Z(\bs{s})$ may be constructed by considering a collection of independent stationary max-stable random fields $Z^1(\bs{s}),\ldots,Z^k(\bs{s})$ with unit Fr\'echet margins and associated proportions $a^1(\bs{s}),\ldots,a^k(\bs{s})\in[0,1]$ such that $\sum_{i=1}^k a^i(\bs{s})=1$ for each $\bs{s}$, yielding the simple max-stable process $Z(\bs{s})=\max_{i=1,\ldots,k}\{a^i(\bs{s})Z^i(\bs{s})\}$. In practice, however, this model may involve too many parameters. 

The second type of mixture consists of sum-mixtures of Gaussian processes \citep{Fuentes:2001,Reich.etal:2011b} used in the formulation of the extremal $t$ model. Specifically, let $\varepsilon^1(\bs{s}),\varepsilon^2(\bs{s})$ be two Gaussian processes with zero mean, unit variance and correlation functions $\rho^1(\bs{s}_1,\bs{s}_2),\rho^2(\bs{s}_1,\bs{s}_2)$, respectively, and let $0\leq a(\bs{s})\leq 1$ be a function defined on $\calS$. Then, a non-stationary extremal $t$ model may be obtained by considering the process $\varepsilon(\bs{s})= a(\bs{s})\varepsilon^1(\bs{s})+\{1-a(\bs{s})\}\varepsilon^2(\bs{s})$ in the construction \eqref{ExtremalT} with correlation function 
\begin{equation}\label{cor.summix}
\rho(\bs{s}_1,\bs{s}_2)={a(\bs{s}_1)a(\bs{s}_2)\rho^1(\bs{s}_1,\bs{s}_2) + \{1-a(\bs{s}_1)\}\{1-a(\bs{s}_2)\}\rho^2(\bs{s}_1,\bs{s}_2)\over [a(\bs{s}_1)^2+\{(1-a(\bs{s}_1)\}^2]^{1/2}[a(\bs{s}_2)^2+\{(1-a(\bs{s}_2)\}^2]^{1/2}}.
\end{equation}
Again, the proportion $a(\bs{s})$ may be modeled in terms of covariates. Similarly, different smoothness behaviors over the space may be captured by the different mixture components. As above, model \eqref{cor.summix} can easily be extended to higher-dimensional mixtures, though this may lead to heavy parametrization. Although similar, the two types of max-stable mixtures are not equivalent, as their corresponding exponent measures differ. 

%%%%%%%%%%%%%%%%%%%%%%%%%%%%%%%%%
%%%%%%%%%%%%%%%%%%%%%%%%%%%%%%%%%
%%%%%%%%%%%%%%%%%%%%%%%%%%%%%%%%%

\section{Inference} \label{sec.inf}

%%%%%%%%%%%%%%%%%%%%%%%%%%%%%%%%%
%%%%%%%%%%%%%%%%%%%%%%%%%%%%%%%%%
\subsection{Pairwise likelihood} \label{sec.pair}
Likelihood inference for max-stable processes is not an easy task. The joint density for max-stable processes stems from the differentiation of (\ref{jointD}) with respect to $z_1,\ldots,z_D$. In dimension $D=2$, this equals $(V_1V_2-V_{12})\exp(-V)$, where $V_1=\partial V_\calD(z_1,z_2)/\partial z_1$, etc., where the subscript $\calD$ and the arguments are dropped for clarity. However, as $D$ increases, the size of this expression renders the full likelihood quickly intractable. To illustrate this, the number of terms in the full likelihood when $D=10,20,50,100$ is of the order $10^5,10^{13},10^{47},10^{115}$, respectively. To get around this computational bottleneck, the use of pairwise likelihoods is now a common practice \citep[see, e.g.,][]{Padoan.etal:2010}. Denoting the vector of unknown parameters by $\bs{\psi}\in\Psi\subset\Real^p$, log pairwise likelihoods for model (\ref{jointD}) may be expressed as
\begin{equation}\label{pairlik}
\ell(\bs{\psi})=\sum_{i=1}^{m} \sum_{(j_1,j_2)\in \calP} \log\left\{V_1(z_{i;j_1},z_{i;j_2})V_2(z_{i;j_1},z_{i;j_2})-V_{12}(z_{i;j_1},z_{i;j_2})\right\} - V(z_{i;j_1},z_{i;j_2}),
\end{equation}
where $z_{i;j}$ denotes the \ith block maximum recorded at the \jth station, $i=1,\ldots,m$, $j=1,\ldots,D$, and where the non-empty set $\calP\subset\calP_{\rm tot}=\{(j_1,j_2):1\leq j_1<j_2\leq D\}$ defines the pairs of observations included in the pairwise likelihood. If $\calP=\calP_{\rm tot}$, all pairs are considered in (\ref{pairlik}). Computational and statistical efficiency might however be gained by carefully selecting a much smaller number of pairs \citep{Huser.Davison:2014a,Castruccio.etal:2016}. A possibility is to include a small fraction of informative pairs, i.e., typically the most dependent ones, though \citet{Huser.Davison:2014a} show that further improvements may be obtained in special cases by including some weakly dependent pairs as well. For stationary isotropic processes, this might be achieved by including the closest pairs, whereas for non-stationary max-stable processes, one might consider pairs $(j_1,j_2)$ with the lowest extremal coefficients $\theta(\bs{s}_{j_1},\bs{s}_{j_2})$. Since the latter are unknown in practice, the choice of pairs might be guided by pre-computed empirical extremal coefficients $\hat\theta(\bs{s}_{j_1},\bs{s}_{j_2})$; however, simulations (not shown) reveal that this approach creates bias, as data are used twice: to select the pairs in the pairwise likelihood and to estimate the parameters by maximizing the latter. Under temporal independence, the maximum pairwise likelihood estimator $\hat{\bs{\psi}}$ maximizing (\ref{pairlik}) is strongly consistent, asymptotically Gaussian, converges at $m^{1/2}$ rate, and its asymptotic variance is of the \emph{sandwich} form, as is typical for mis-specified likelihood estimators \citep{Padoan.etal:2010}. More precisely, if $\bs{\psi}_0\in{\rm int}(\Psi)$ denotes the ``true'' parameter vector, then under mild regularity conditions, one has the large sample approximation
\begin{equation}\label{asymp}
\hat{\bs{\psi}}\dotsim\calN_p(\bs{\psi}_0,\bs{J}(\bs{\psi}_0)^{-1}\bs{K}(\bs{\psi}_0)\bs{J}(\bs{\psi}_0)^{-1}),\quad m\to\infty,
\end{equation}
where $\bs{J}(\bs{\psi})=\E\{-\partial^2\ell(\bs{\psi})/\partial\bs{\psi}\partial\bs{\psi}^T\}\in\Real^{p\times p}$ and $\bs{K}(\bs{\psi})=\var\{\partial\ell(\bs{\psi})/\partial\bs{\psi}\}\in\Real^{p\times p}$. Uncertainty may be assessed by plugging estimates of the matrices $\bs{J}(\bs{\psi}_0)$ and $\bs{K}(\bs{\psi}_0)$ into the asymptotic variance in (\ref{asymp}); see \citet{Padoan.etal:2010}. Alternatively, one can bootstrap the independent replicates $\bs{z}_i=(z_{i;1},\ldots,z_{i;D})^T$, $i=1,\ldots,m$, and re-estimate parameters using the pseudo-samples, to assess the variability surrounding $\hat{\bs{\psi}}$. Similar asymptotic properties hold for mildly time-dependent processes \citep{Davis.etal:2013b,Huser.Davison:2014a} in which uncertainty may be assessed using block bootstrap. 

%%%%%%%%%%%%%%%%%%%%%%%%%%%%%%%%%
%%%%%%%%%%%%%%%%%%%%%%%%%%%%%%%%%
\subsection{Goodness-of-fit assessment and model selection} \label{sec.good}
Model comparison is typically performed using the composite likelihood information criterion (CLIC), defined as ${\rm CLIC}=-2\ell(\hat{\bs{\psi}}) + 2{\rm tr}\{\bs{J}(\hat{\bs{\psi}})^{-1}\bs{K}(\hat{\bs{\psi}})\}$, which is comparable to the Akaike information criterion. Another possibility is to use the composite Bayesian information criterion (CBIC), i.e., the counterpart of the classical Bayesian information criterion. It is defined as ${\rm CBIC}=-2\ell(\hat{\bs{\psi}}) + \log(m){\rm tr}\{\bs{J}(\hat{\bs{\psi}})^{-1}\bs{K}(\hat{\bs{\psi}})\}$, and therefore penalizes model complexity more than does CLIC. The lower the CLIC or CBIC, the better the model. Theoretical properties of CLIC and CBIC have been investigated by \citet{Ng.Joe:2014} (in which CLIC and CBIC are called instead CLAIC and CLBIC, respectively). In particular, they show that CLIC has a tendency to select over-complicated models. For a broad survey of composite likelihood methods, see \citet{Varin.etal:2011}.

%%%%%%%%%%%%%%%%%%%%%%%%%%%%%%%%%
%%%%%%%%%%%%%%%%%%%%%%%%%%%%%%%%%
%%%%%%%%%%%%%%%%%%%%%%%%%%%%%%%%%
\section{Simulation study} \label{sec.sim}
%%%%%%%%%%%%%%%%%%%%%%%%%%%%%%%%%
%%%%%%%%%%%%%%%%%%%%%%%%%%%%%%%%%
\subsection{Setup}
In this simulation study, we assess the ability of the maximum pairwise likelihood estimator (\ref{asymp}) to estimate and detect non-stationarity dependence structures in a variety of contexts. We also study the effect of neglecting non-stationarity on spatial return levels.

Throughout this section, we focus on the \emph{locally isotropic} extremal $t$ model illustrated in the first row of Figure~\ref{Fig:NonStatExtremalT} and consider various parameter combinations. Specifically, the extremal $t$ process with ${\rm df}=1,2,5,10$ is simulated on $[0,1]^2$, using the non-stationary correlation function $\rho(\bs{s}_1,\bs{s}_2)$ defined in (\ref{PaciorekCorrelation}) based on the powered exponential model (\ref{PoweredExponential}) with $\alpha=0.5,1,1.5,1.9$ (rough to smooth). The underlying spatially varying covariance matrix is taken to be of the form $\bs{\Omega}_{\bs{s}}=(2\,{\rm df})^{2/\alpha}\times\omega({\bs{s}})^2\bs{I}_2$, where $\bs{I}_2$ is the $2$-by-$2$ identity matrix and $\omega({\bs{s}})=\beta_1 2^{-\beta_2|s_x|}$, $\bs{s}=(s_x,s_y)$, with range $\beta_1>0$ and non-stationary parameter $\beta_2\in\Real$. To investigate different non-stationary scenarios, we consider $(\beta_1,\beta_2)=(0.1,0)$ (stationary), $(0.1\sqrt{2},1)$ (weakly non-stationary), $(0.2,2)$ (mildly non-stationary), and $(0.4,4)$ (strongly non-stationary). Although these scenarios exhibit different non-stationarity patterns, the overall dependence strength is comparable in the sense that all cases satisfy $\omega({\bs{s}})=0.1$ for any $\bs{s}=(0.5,s_y)$. The ${\rm df}=1$ case corresponds to a non-stationary Schlather process, whereas the ${\rm df}=10$ case is a crude approximation of a non-stationary Brown--Resnick process (with $\alpha=1.9$ corresponding approximately to the non-stationary Smith model); recall Section~\ref{sec.ext}. In each case, $m=10,20,50,100$ independent replicates of these processes are simulated at $S=10,20,50,100$ fixed locations uniformly sampled in the unit square. Simulations are repeated $300$ times to compute empirical diagnostics.

\subsection{Estimation and detection of non-stationarity} 
We first investigate the performance of the maximum pairwise likelihood estimator (\ref{asymp}) to recover the true parameters under the correct model. We estimate parameters $\bs{\psi}=(\beta_1,\beta_2,{\rm df},\alpha)^T$ with (\ref{asymp}) using the 10\% closest pairs; then we derive the empirical biases, standard deviations and root mean squared errors (RMSE) from the $300$ independent experiments. RMSEs, typically dominated by the standard deviations, are reported in Table~\ref{TableResults1}.

We focus on the estimation of $\beta_1$ and $\beta_2$, which determine the non-stationary scenario. The range parameter $\beta_1$ is quite well identified overall. The corresponding RMSE is less than $0.02$, $0.04$, $0.06$ and $0.12$ for $\beta_1=0.1$, $0.1\sqrt{2}$, $0.2$, and $0.4$, respectively, and it decreases as the smoothness parameter $\alpha$ increases, and as the degrees of freedom (${\rm df}$) increase. Furthermore, the higher $\beta_1$, the larger its RMSE, as expected. The RMSE for the non-stationary parameter $\beta_2$ follows a similar pattern, though large values of $\beta_2$ seem easier to estimate overall: for strongly non-stationary scenarios, the RMSE is quite small in comparison to the actual value of $\beta_2$. This is certainly due to the very rigid type of assumed non-stationarity: a small perturbation of $\beta_2$ entails a dramatic change in the dependence structure. 
%% Figure 4
%\begin{figure}[t!]
%\centering
%\includegraphics[width=0.85\linewidth]{Boxplots_bis.pdf}
%\caption{Boxplots of parameter estimates obtained from data generated from the locally isotropic extremal $t$ process with ${\rm df}=5$, $\alpha=1$ and $(\beta_1,\beta_2)=(0.2,2)$. Estimator (\ref{asymp}) was used, including the 10\% closest pairs. Green boxes (left of vertical dashed line) show the performance for a fixed number of locations, $S=100$, and an increasing number of independent replicates, $m=10,20,50,100$. Blue boxes (right of vertical dashed line) show the performance for a fixed number of replicates, $m=100$, and an increasing number of locations, $S=10,20,50,100$. Horizontal red lines are true values.}\label{Fig:Boxplots}
%\end{figure}

\begin{figure}[t!]
\centering
\includegraphics[width=0.85\linewidth]{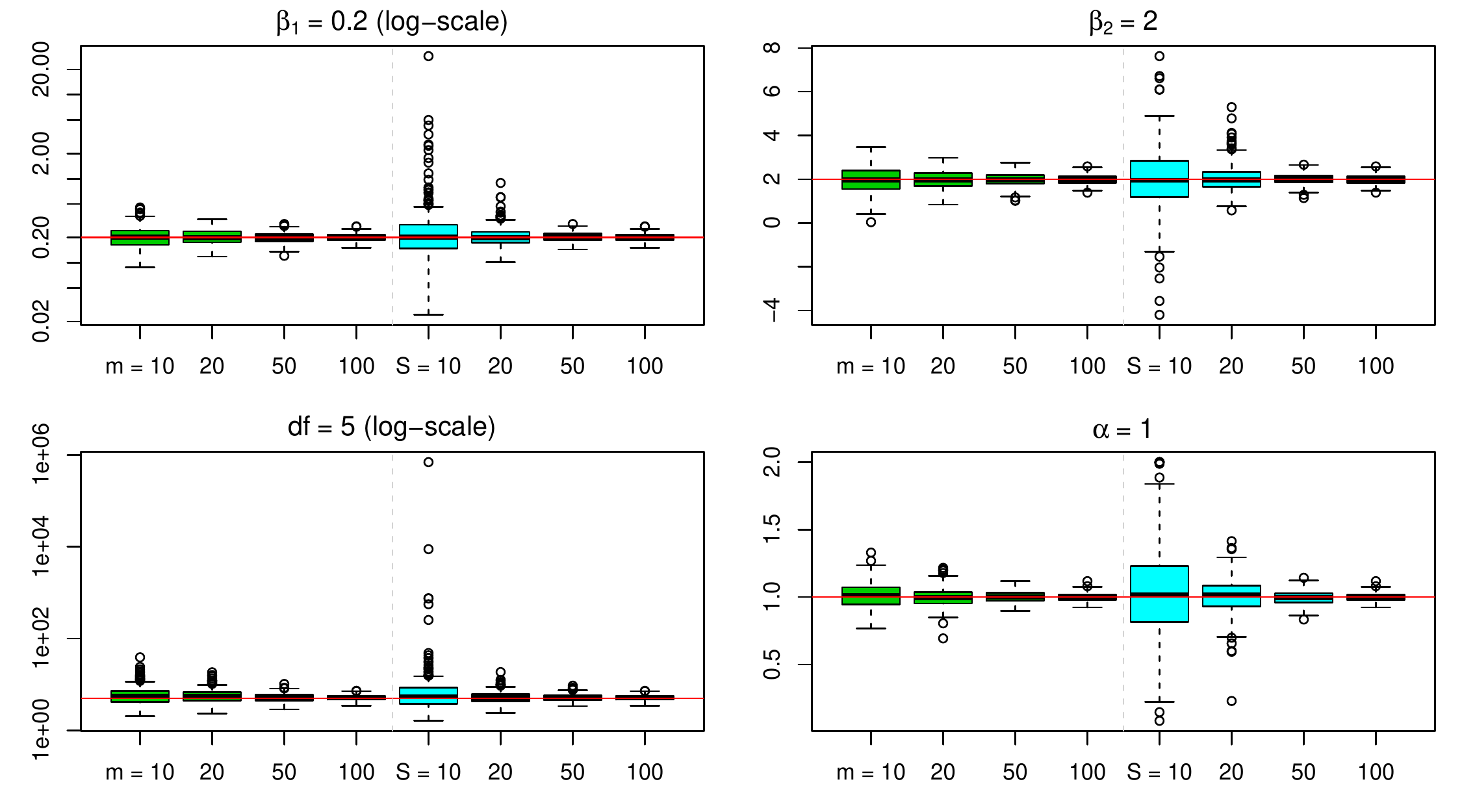}
\caption{Boxplots of parameter estimates obtained from data generated from the locally isotropic extremal $t$ process with ${\rm df}=5$, $\alpha=1$ and $(\beta_1,\beta_2)=(0.2,2)$. Estimator (\ref{asymp}) was used, including the 10\% closest pairs. Green boxes (left of vertical dashed line) show the performance for a fixed number of locations, $S=100$, and an increasing number of independent replicates, $m=10,20,50,100$. Blue boxes (right of vertical dashed line) show the performance for a fixed number of replicates, $m=100$, and an increasing number of locations, $S=10,20,50,100$. Horizontal red lines are true values.}\label{Fig:Boxplots}
\end{figure}

To illustrate increasing-domain and infill asymptotic properties of the estimator (\ref{asymp}), Figure~\ref{Fig:Boxplots} displays boxplots of parameter estimates, as a function of $m$ and $S$ for the extremal $t$ model with ${\rm df}=5$, $\alpha=1$ and $(\beta_1,\beta_2)=(0.2,2)$. As expected, the estimator appears to be consistent as $m$ increases. In addition, parameters are much better estimated if the data are collected at a dense network of sites, although the estimator is not consistent as $S\to\infty$ for fixed $m$, as a result of the extremal-$t$ model being non-mixing. Interestingly, the estimated variances of $\beta_1/\beta_2/{\rm df}/\alpha$ decrease by a factor $4.9/4.7/8.3/4.8$ when the number of \emph{independent} repeated measured increases from $m=20$ to $m=100$ (for $S=100$), whereas they drop by a factor $13.2/9.9/5.8/17.2$ when the number of \emph{dependent} spatial measurements increases from $S=20$ to $S=100$ (for $m=100$). Therefore, in finite samples, having more stations may be (much) more valuable than having more replicates.

We now explore the ability of estimator (\ref{asymp}) to detect the spatial heterogeneity. For each simulated dataset, we fit the true non-stationary model and the (restricted) stationary counterpart, computing in each case the corresponding CLIC and CBIC diagnostics defined in Section~\ref{sec.pair}. These information criteria were computed using finite differences combined with the direct method of \citet{Padoan.etal:2010}. The empirical percentages that the CLIC and CBIC are in favor of the true underlying model (either stationary if $\beta_2=0$, or non-stationary otherwise) are calculated from the $300$ experiments and reported in the Supplementary Material for $S=100$ and $m=100$. Overall, non-stationarity in the dependence structure seems easily detectable when the non-stationarity level is moderate to strong, with almost $100\%$ of success in each case with the CLIC or CBIC. By contrast, the performance is poor in near-stationary cases; this is especially striking for the CBIC, which penalizes more model complexity. In case of stationarity, the CLIC selects the true model in about $65\%$ of occasions, whereas the CBIC attains about $80\%$ of success. This suggests that these information criteria, but especially the CLIC, have ``more power'' to select bigger models, and that they should be interpreted with care. This observation agrees with the theoretical findings of \citet{Ng.Joe:2014}. Furthermore, the ability to distinguish between stationarity and non-stationarity improves when more data are available. 
For example, for fixed $S=20$ and parameters ${\rm df}=5$, $\alpha=1$, $(\beta_1,\beta_2)=(0.2,2)$,  the CLIC percentages are $63\%,79\%,93\%,99\%$, for $m=10,20,50,100$, respectively; similarly, for fixed $m=20$, these values are $42\%,79\%,98\%,100\%$, for $S=10,20,50,100$, respectively.

\subsection{Effect of model misspecification on return levels} 
Neglecting non-stationarity when the data are in fact non-stationary might have serious consequences on the estimation of spatial return levels. To assess this, we consider the locally isotropic extremal $t$ model on the Gumbel scale, with ${\rm df}=5$ and $\alpha=1.5$. For $(\beta_1,\beta_2)=(0.1,0)$ (stationary case) and $(\beta_1,\beta_2)=(0.4,4)$ (strongly non-stationary case), we compute return levels for the integral ${\rm INT}_j=\int_{\calS_j} Z(\bs{s}) {\rm d}\bs{s}$, the minimum  ${\rm MIN}_j=\min_{\bs{s}\in{\calS_j}} \{Z(\bs{s})\}$, and the maximum ${\rm MAX}_j=\max_{\bs{s}\in{\calS_j}} \{Z(\bs{s})\}$, $j=1,2$, of the max-stable process $Z(\bs{s})$ over the domains $\calS_1=[0,0.2]\times[0,1]$ and $\calS_2=[0.8,1]\times[0,1]$. In practice, these domains are pixelated using a fine grid comprising $105$ points with equal spacings of $0.05$. Assuming that $Z(\bs{s})$ describes the annual maximum process for some quantity of interest, we then derive the $N$-year return level for ${\rm INT}_j$ and ${\rm MIN}_j$ as the empirical $(1-1/N)$-quantile calculated from one million independent simulations of $Z(\bs{s})$. Return levels $z_{N;{\rm MAX}_j}$ for ${\rm MAX}_j$ are derived using the exact formula $z_{N;{\rm MAX}_j}=\log\{\theta(\calS_j)\}-\log\{-\log(1-1/N)\}$ and an estimate of the areal extremal coefficient $\theta(\calS_j)$ \citep{Lantuejoul.etal:2011}. The latter determines the effective number of independent extremes in region $\calS_j$; for the stationary case, one finds $\theta(\calS_1)=\theta(\calS_2)\approx8.6$, and for the non-stationary case, $\theta(\calS_1)\approx4.2$, $\theta(\calS_2)\approx23.6$, indicating that extremal dependence in $\calS_1$ is much stronger than in $\calS_2$. Results are shown in Figure~\ref{Fig:ReturnLevels}.

One can see that mis-specification (and therefore also mis-estimation) of spatial dependence strongly affects the return levels of spatial quantities. Underestimation of dependence implies underestimation of return levels for ${\rm INT}_j$ and ${\rm MIN}_j$ and overestimation of return levels for ${\rm MAX}_j$ (and vice versa). Although this depends on the level of non-stationarity, the underlying parameters, and marginal distributions, in practice it is crucial to capture correctly the non-stationarity in the dependence structure.

%%%%%%%%%%%%%%%%%%%%%%%%%%%%%%%%%
%%%%%%%%%%%%%%%%%%%%%%%%%%%%%%%%%
%%%%%%%%%%%%%%%%%%%%%%%%%%%%%%%%%

\section{Analysis of temperature maxima} \label{sec.ill}
\begin{figure}[t!]
\centering
\begin{subfigure}{0.4\linewidth}
\includegraphics[width=\linewidth]{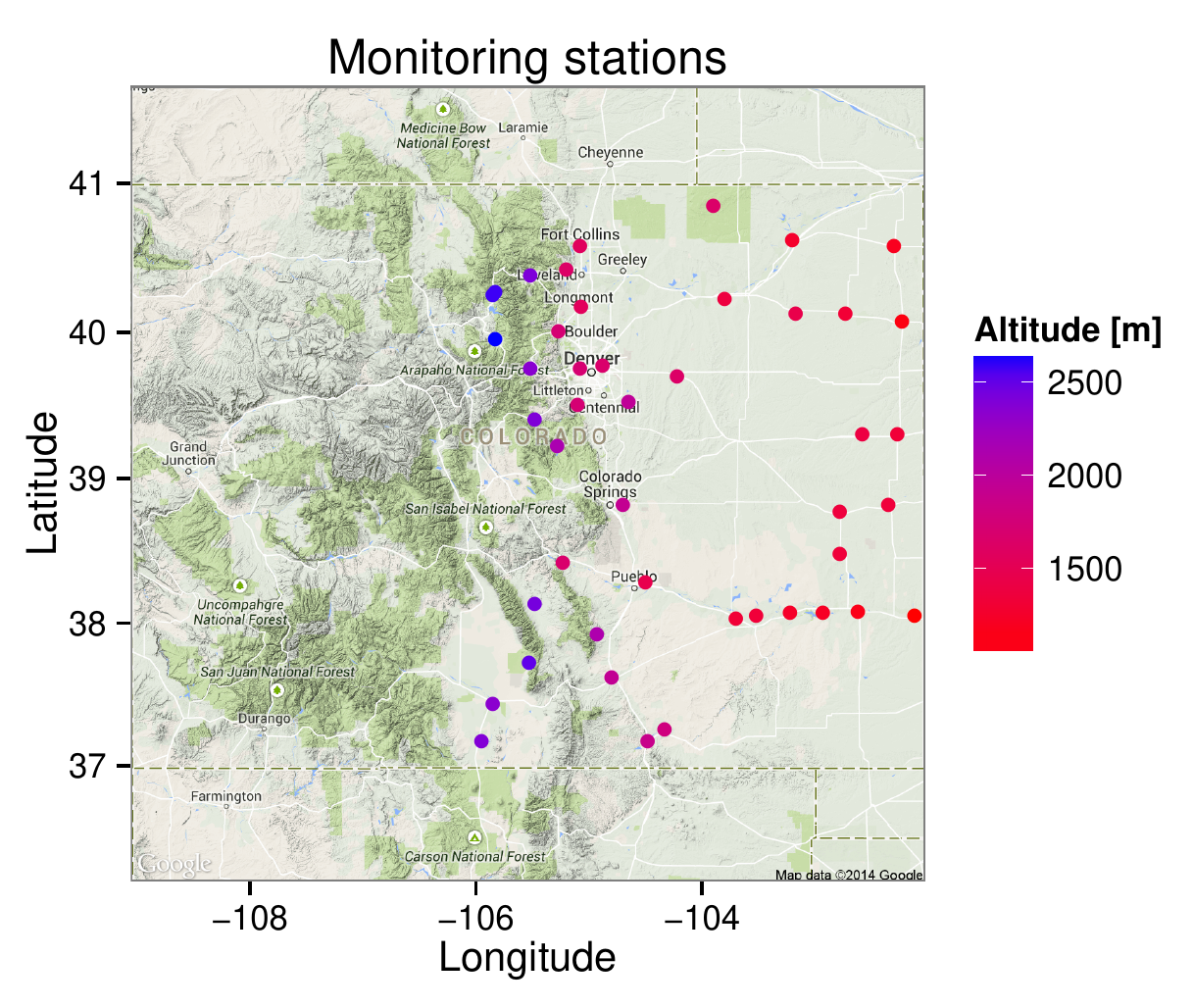}
\end{subfigure}
\begin{subfigure}{0.59\linewidth}
\includegraphics[width=\linewidth]{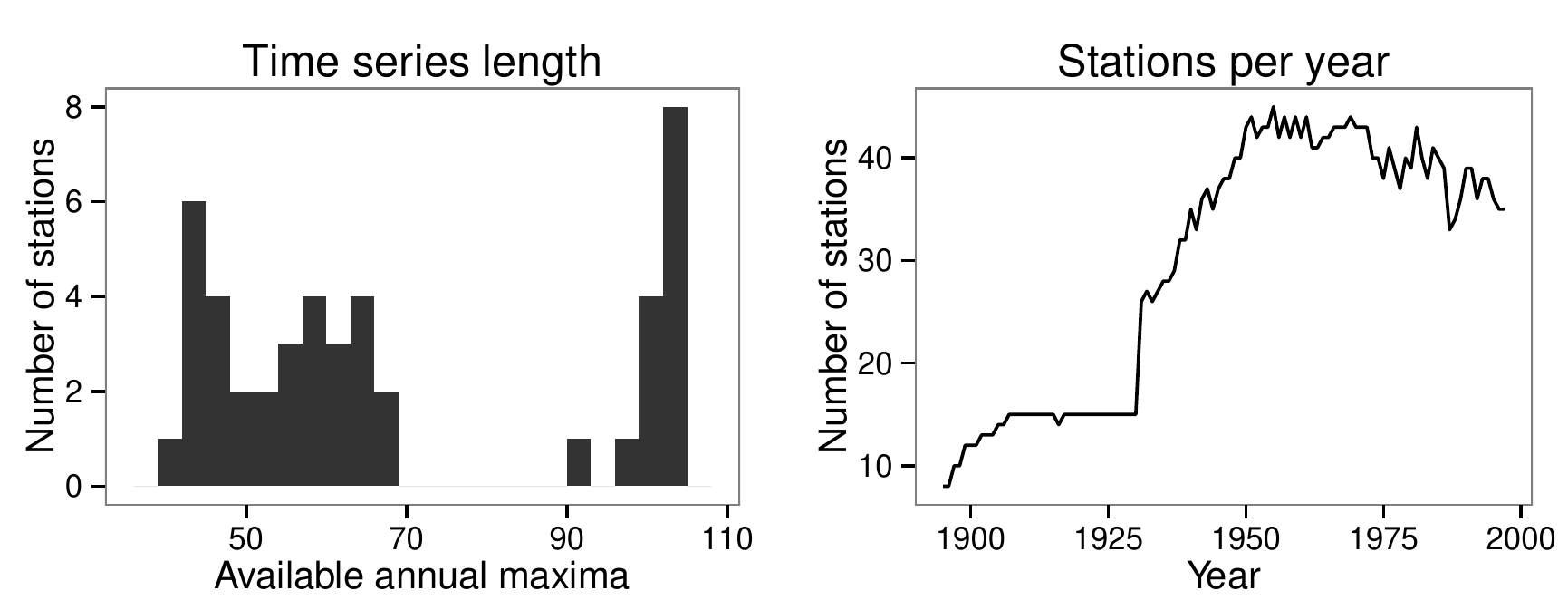}
\end{subfigure}
\caption{\emph{Left}: Map of Colorado with the $45$ stations (dots) used in the analysis of temperature maxima. \emph{Middle}: a histogram summarizing the number of annual maxima available per station. \emph{Right}: Number of stations per year.}\label{Fig:Illustration}
\end{figure}

We now discuss an application to a temperature dataset recorded in Colorado during the period $1895$-$1997$, which is freely available on the National Center for Atmospheric Research website. We selected stations in the Front Range area, with at least $40$ years of data, and extracted maxima over the months May--September (roughly corresponding to annual maxima), bypassing therefore the modeling of seasonality. Figure~\ref{Fig:Illustration} illustrates the locations of the monitoring stations kept for the analysis, and summarizes the data availability. 

%%Figure 5
%\begin{figure}[t!]
%\centering
%\begin{subfigure}{0.4\linewidth}
%\includegraphics[width=\linewidth]{Illustration_a.pdf}
%\end{subfigure}
%\begin{subfigure}{0.59\linewidth}
%\includegraphics[width=\linewidth]{Illustration_b.pdf}
%\end{subfigure}
%\caption{\emph{Left}: Map of Colorado with the $45$ stations (dots) used in the analysis of temperature maxima. \emph{Middle}: a histogram summarizing the number of annual maxima available per station. \emph{Right}: Number of stations per year.}\label{Fig:Illustration}
%\end{figure}

\begin{figure}[t!]
\centering
\includegraphics[width=0.85\linewidth]{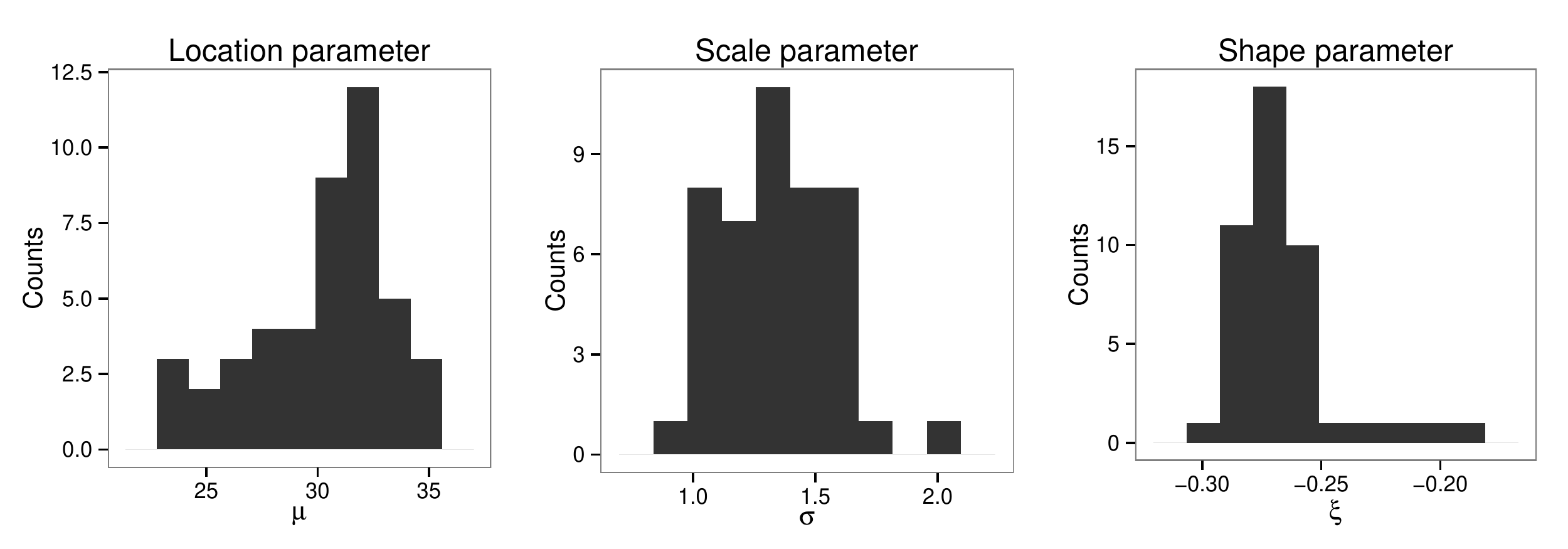}
\caption{Histograms of estimated location $\mu(\bs{s})$ (left), scale $\sigma(\bs{s})$ (middle) and shape $\xi(\bs{s})$ (right) parameters obtained from the fit of the spatial GEV$(\mu(\bs{s}),\sigma(\bs{s}),\xi(\bs{s}))$ model with parameters modeled as (conditionally independent) latent Gaussian processes.}\label{Fig:MarginalFits}
\end{figure}

To estimate marginal distributions, we fitted a spatial GEV$(\mu(\bs{s}),\sigma(\bs{s}),\xi(\bs{s}))$ model to observed maxima, assuming conditional independence with the parameters $\mu(\bs{s}),\sigma(\bs{s}),\xi(\bs{s})$, modeled as latent stationary Gaussian processes. While the means of the location and scale parameters $\mu(\bs{s}),\sigma(\bs{s})$ were assumed to depend on longitude, latitude and altitude, the mean of the shape parameter involved only two distinct values for plains and mountains. Quantile-quantile plots (not shown) suggest that marginal fits are good. Annual maxima were then transformed to the unit Fr\'echet scale using the parameters' estimated mean and the probability integral transform. Histograms of estimated parameters for the different stations are displayed in Figure~\ref{Fig:MarginalFits}. Shape parameters are all negative, indicating that distributions of temperature annual maxima have an upper bound, which seems physically plausible. 

We then fitted 16 stationary and non-stationary extremal $t$ models to the transformed data using the pairwise likelihood estimator \eqref{asymp} including all pairs of locations. These models, summarized in Table~\ref{MSmodels}, are based on the Paciorek--Schervish correlation function \eqref{PaciorekCorrelation} combined with \eqref{PoweredExponential} and are parametrized as in \eqref{MatrixOmega}. 
They are either stationary (models $1$--$2$) or non-stationary (models $3$--$16$), locally isotropic (models $1,3$--$5,9,11$--$13$) or anisotropic (models $2,6$--$8,10,14$--$16$), based on Gaussian sum-mixtures of the form \eqref{cor.summix} (models $1$--$8$) or non-mixtures (models $9$--$16$). In the non-stationary models, altitude, longitude and latitude are used as covariates (on top of the intercept) in the modeling of the dependence ranges $\omega_x(\bs{s}),\omega_y(\bs{s})$ (with logarithmic link) and the mixture coefficient $a(\bs{s})$ (with logit link), as suggested in \eqref{CovariatesModeling} and Section~\ref{sec.alt}. The anisotropy parameter $\delta(\bs{s})$ is kept constant. The degrees of freedom, ${\rm df}$, were found to be difficult to estimate, and after some analysis, were held fixed at ${\rm df}=5$ (i.e., far from the Smith--Stephenson and Brown--Resnick families).

\begin{figure}[t!]
\centering
\includegraphics[width=0.85\linewidth]{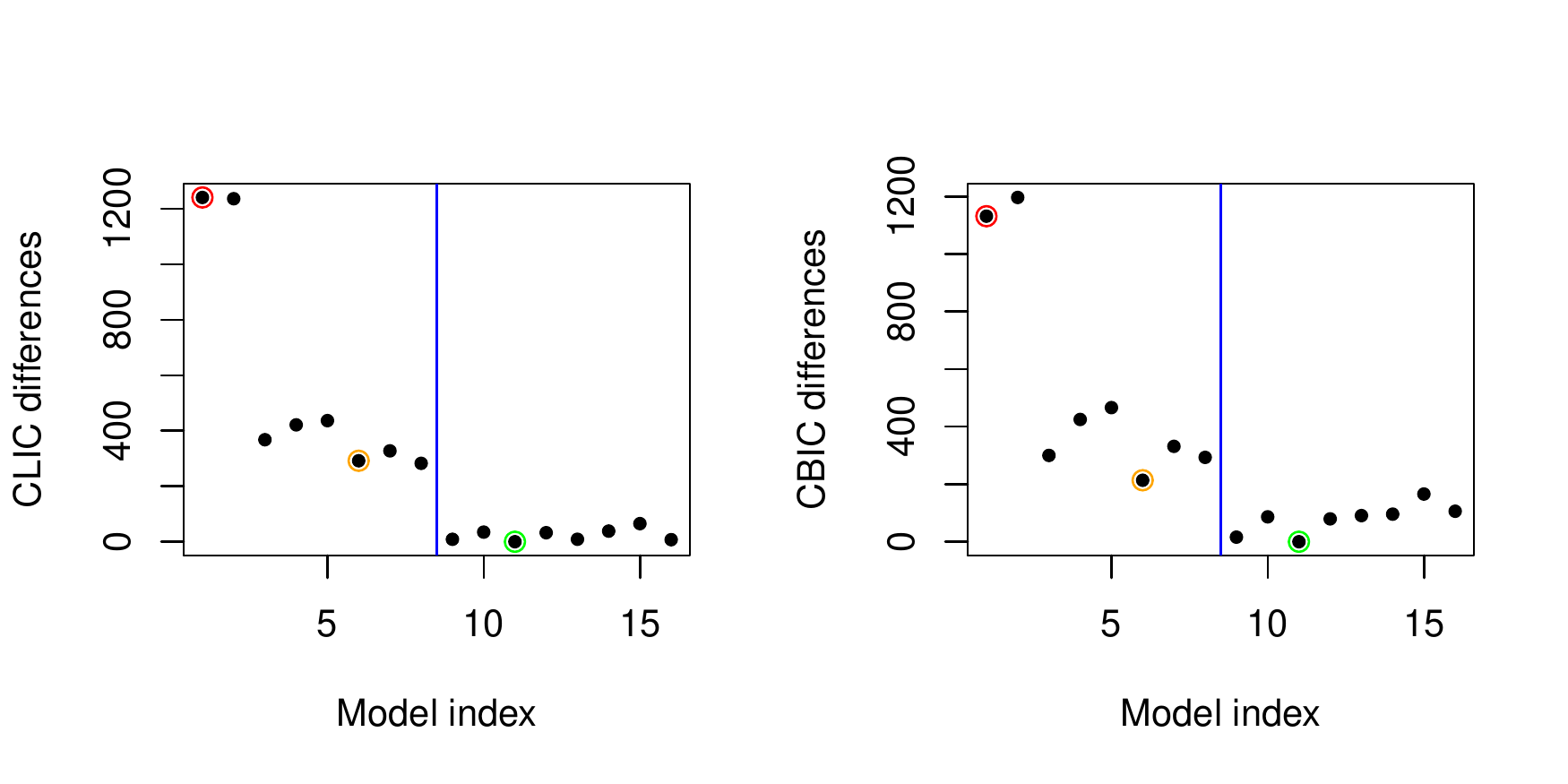}
\caption{Difference of estimated CLIC (left) and CBIC (right) values for all max-stable models fitted with respect to the best fit. Vertical blue lines mark the separation between mixture ($1$--$8$) and non-mixture ($9$--$16$) models. Models used for comparison are highlighted in red (stationary isotropic model, 1), orange (best non-mixture model, 6) and green (best mixture model, 11)}\label{Fig:CLICS_CBICS}
\end{figure}

Figure~\ref{Fig:CLICS_CBICS} reports the estimated CLIC and CBIC values of the fitted models; recall Section~\ref{sec.pair}. These two diagnostics agree on at least two main conclusions:
\begin{enumerate}[(i)]
\item \textbf{Mixture models fit generally better}, although they have three more parameters than their non-mixture counterparts. The rougher mixture component tends to be dominant in the mountainous region, while the smoother one (though not very smooth) takes over at lower altitudes.
\item \textbf{Altitude is a major covariate} to be considered in the modeling of extremal dependence, whereas inclusion of further covariates (longitude or latitude) does not improve the fit by much. In non-mixture models, there is a huge drop in CLIC or CBIC values between model $1$ (stationary isotropic model with two parameters) and model $3$ (locally isotropic model, including altitude as a covariate, with only three parameters). In mixture models, point (i) underscores the importance of having different degrees of regularity at different altitudes.
\end{enumerate}

%% Figure 7
%\begin{figure}[t!]
%\centering
%\includegraphics[width=0.85\linewidth]{CLICS_CBICS.pdf}
%\caption{Difference of estimated CLIC (left) and CBIC (right) values for all max-stable models fitted with respect to the best fit. Vertical blue lines mark the separation between mixture ($1$--$8$) and non-mixture ($9$--$16$) models. Models used for comparison are highlighted in red (stationary isotropic model, 1), orange (best non-mixture model, 6) and green (best mixture model, 11)}\label{Fig:CLICS_CBICS}
%\end{figure}

Among non-mixture models, it is worth considering non-stationary non-isotropic models with covariates included in the dependence ranges $\omega_x(\bs{s})$, $\omega_y(\bs{s})$. The best non-mixture model is model $8$ (respectively $6$) according to the CLIC (respectively CBIC), but CLIC tends to select overcomplicated models. For mixture models with altitude included in the mixture coefficient $a(\bs{s})$, use of further covariates in $\omega_x(\bs{s})$, $\omega_y(\bs{s})$ does not improve the fit by much, although both diagnostics agree to select model 11 as the best model.

\begin{figure}[t!]
\centering
\includegraphics[width=0.85\linewidth]{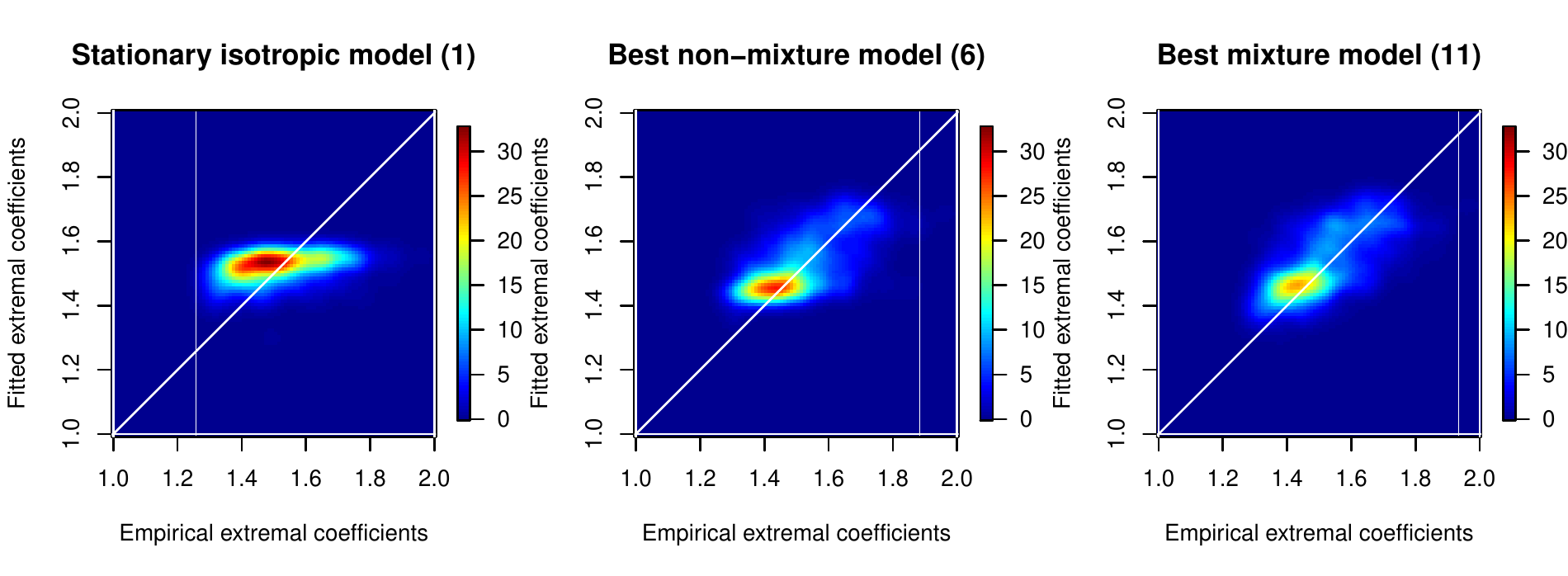}
\caption{Bivariate kernel density estimators for pairs of empirical and fitted extremal coefficients, displayed for (left) model $1$ (stationary isotropic extremal $t$ model), (middle) model $6$ (best non-mixture model) and (right) model $11$ (best mixture model). A good fit should have points concentrated around the white diagonal line.}\label{Fig:Extremal_Coefs}
\end{figure}

Figure~\ref{Fig:Extremal_Coefs} displays bivariate kernel density estimators for the pairs of empirical and fitted extremal coefficients for model $1$ (stationary isotropic benchmark), model 6 (best non-mixture model according to the CBIC) and model $11$ (best mixture model). Empirical estimates are calculated using the projection method of \citet{Marcon.etal:2014} based on the non-parametric Pickands dependence estimator of \citet{Caperaa.etal:1997}.
%% Figure 8
%\begin{figure}[t!]
%\centering
%\includegraphics[width=0.85\linewidth]{ExtremalCoefficients.pdf}
%\caption{Bivariate kernel density estimators for pairs of empirical and fitted extremal coefficients, displayed for (left) model $1$ (stationary isotropic extremal $t$ model), (middle) model $6$ (best non-mixture model) and (right) model $11$ (best mixture model). A good fit should have points concentrated around the white diagonal line.}\label{Fig:Extremal_Coefs}
%\end{figure}
Extremal dependence is slightly underestimated for model $1$ (with a majority of points lying above the diagonal line), but extremal coefficients for non-stationary models tend to be generally closer to the diagonal. The sum of squared distances between fitted and empirical extremal coefficients is $3.63,3.12,3.00$ for models $1,6,11$, respectively. Clearly, the stationary isotropic model provides the worse fit, which confirms our previous conclusions, and even more strongly supports the need for non-stationary dependence structures to incorporate meaningful covariates.

%%%%%%%%%%%%%%%%%%%%%%%%%%%%%%%%%
%%%%%%%%%%%%%%%%%%%%%%%%%%%%%%%%%
%%%%%%%%%%%%%%%%%%%%%%%%%%%%%%%%%

\section{Discussion} \label{sec.dis}
The problem of building and fitting sensible non-stationary dependence models for spatial extremes is not trivial. We have tackled this problem by proposing a very general construction, combining max-stable processes (in particular the extremal $t$ model), non-stationary correlation functions, and mixtures. The advocated locally elliptic model is based on \citet{Paciorek.Schervish:2006} and allows various non-stationary patterns to be flexibly captured in the extremal dependence structure by incorporating meaningful covariates. We have performed inference using pairwise likelihoods, which are computationally convenient, and we have shown by simulation that pairwise likelihoods can efficiently estimate the unknown parameters, provided that the station network is dense. However, more efficient approaches based on full likelihoods \citep{Stephenson.Tawn:2005,Wadsworth.Tawn:2014,Thibaud.Opitz:2015} might be devised for the extremal $t$ model. 

Various non-stationary max-stable models, including altitude, longitude and latitude as covariates, were fitted to a dataset of temperature maxima in Colorado, and these models were shown to provide a better fit with respect to the traditional stationary and isotropic max-stable counterpart, although there is still room for improvement. In particular, we have identified altitude as an important covariate.  
In future work, other covariates, such as the slope or solar radiation, might be used to improve the fit, perhaps from satellite data or regional climate computer models. Alternatively, more flexible non-stationary models might be constructed from a Bayesian perspective, though inference may be tricky and computationally very intensive if standard Markov chain Monte Carlo algorithms are used \citep[but see][]{Thibaud.etal:2015}. The creation of models for asymptotic independence, a degenerate case in the max-stable paradigm, is also an important issue when data are non-stationary. One possibility could be to ``invert'' the non-stationary max-stable models proposed above \citep[see][]{Wadsworth.Tawn:2012b,Davison.etal:2013}.

Finally, we focused in this work on maxima, but more efficient approaches may be achieved by considering peaks over high thresholds \citep{Huser.etal:2016}. This approach, however, entails additional complications such as the modeling of temporal dependence, the selection of a suitable threshold and the non-validity of extremal models at low levels, which might be even more difficult to handle when the data are non-stationary.

\baselineskip 15pt

%\bibliographystyle{CUP}
%\bibliography{BigBib}

\baselineskip 10pt

\clearpage
\thispagestyle{empty}

\begin{table}[t!]
\centering
\footnotesize
\caption{Root mean squared error ($\times 100$) of the maximum pairwise likelihood estimator (\ref{asymp}), using the 10\% closest pairs, for the locally isotropic extremal $t$ with various parameter combinations. These diagnostics are computed for parameters $\beta_1/\beta_2/{\rm df}/\alpha$ from $300$ independent simulations of $m=100$ independent max-stable processes simulated at $S=100$ fixed locations in $[0,1]^2$.}\label{TableResults1}
\vspace{3pt}
\begin{tabular}{rr|cccc}
\multicolumn{2}{c|}{}&\multicolumn{4}{c}{Smoothness parameter $\alpha$}\\
$(\beta_1,\beta_2)$ & ${\rm df}$ & $0.5$ & $1.0$ & $1.5$ & $1.9$ \\
\hline
$(0.1,0)$ & $1$ & $2/42/12/3$ & $1/21/11/4$ & $1/14/10/5$ & $1/11/10/5$ \\
 & $2$ & $2/41/30/3$ & $1/21/27/3$ & $1/13/23/5$ & $0/10/22/4$ \\
 & $5$ & $2/32/113/3$ & $1/17/92/3$ & $0/11/92/3$ & $0/10/100/3$ \\
 & $10$ & $1/27/434/2$ & $1/15/302/3$ & $0/11/277/3$ & $0/8/417/3$ \\[3pt]
$(0.1\sqrt{2},1)$ & $1$ & $4/47/12/3$ & $2/22/10/4$ & $1/15/11/5$ & $1/11/10/6$ \\
 & $2$ & $3/43/30/3$ & $1/20/26/3$ & $1/14/22/4$ & $1/10/23/4$ \\
 & $5$ & $3/36/111/3$ & $1/19/93/3$ & $1/12/88/3$ & $1/10/93/3$ \\
 & $10$ & $2/28/285/3$ & $1/17/259/3$ & $1/12/313/3$ & $1/10/328/3$ \\[3pt]
$(0.2,2)$ & $1$ & $6/55/12/4$ & $2/23/11/4$ & $1/15/9/5$ & $1/13/8/5$ \\
 & $2$ & $5/43/29/3$ & $2/21/25/3$ & $1/15/23/4$ & $1/12/26/4$ \\
 & $5$ & $4/31/92/3$ & $2/20/78/3$ & $1/13/80/3$ & $1/11/81/3$ \\
 & $10$ & $3/24/217/3$ & $2/17/183/3$ & $1/13/193/3$ & $1/11/200/3$ \\[3pt]
$(0.4,4)$ & $1$ & $12/57/12/4$ & $7/31/10/5$ & $4/20/8/6$ & $3/16/7/5$ \\
 & $2$ & $8/41/26/4$ & $5/24/23/4$ & $4/18/21/4$ & $3/15/18/4$ \\
 & $5$ & $7/38/75/4$ & $4/20/57/4$ & $3/14/55/4$ & $2/13/52/4$ \\
 & $10$ & $7/37/194/4$ & $4/20/128/3$ & $3/15/128/4$ & $2/12/116/4$
\end{tabular}
\end{table}

\begin{table}[h!]
\centering
\small
\caption{Extremal $t$ max-stable models fitted to the temperature maxima. For each of the these models, we report whether they are stationary (Stat.), locally isotropic (Iso.), and based on Gaussian sum-mixtures (Mix.). If a model is non-stationary, altitude (Alt.), longitude (Lon.) and latitude (Lat.) may be used as covariates in the dependence ranges $\omega_x(\bs{s}),\omega_y(\bs{s})$ and the mixture coefficient $a(\bs{s})$; recall \eqref{MatrixOmega} and \eqref{CovariatesModeling}. The anisotropy parameter $\delta(\bs{s})$ is kept constant. If a model is locally isotropic, $\omega_x(\bs{s})=\omega_y(\bs{s})$ and $\delta(\bs{s})=0$. Mixture models are constructed from two correlation functions of the form \eqref{QuadraticForm} combined with \eqref{PaciorekCorrelation}, with different smoothness parameters $\alpha_1,\alpha_2$, but based on the same matrix $\bs{\Omega}_{\bs{s}}$. The degrees of freedom are fixed to ${\rm df}=5$. The total number of parameters to estimate (Nb. par.) is also reported.}\label{MSmodels}
\vspace{5pt}
\begin{tabular}{c|ccc|cccc|c}
\multicolumn{4}{c|}{} & \multicolumn{4}{c|}{Covariates included in} & \\[3pt]
\# & Stat. & Iso. & Mix. & $\omega_x({\bs{s}})$ & $\omega_y({\bs{s}})$ & $\delta({\bs{s}})$ & $a({\bs{s}})$ & Nb. par.\\
\hline
$1$ & Yes & Yes & No &  none & --- & --- & --- & 2\\
$2$ & Yes & No & No & none & none & none & --- & 4\\
$3$ & No & Yes & No & Alt. & --- & --- & --- & 3\\
$4$ & No & Yes & No & Alt./Lon. & --- & --- & --- & 4\\
$5$ & No & Yes & No & Alt./Lon./Lat. & --- & --- & --- & 5\\
$6$ & No & No & No & Alt. & Alt. & none & --- & 6\\
$7$ & No & No & No & Alt./Lon. & Alt./Lon. & none & --- & 8\\
$8$ & No & No & No & Alt./Lon./Lat. & Alt./Lon./Lat. & none & --- & 10\\
$9$ & No & Yes & Yes &  none & --- & --- & Alt. & 5\\
$10$ & No & No & Yes & none & none & none & Alt. & 7\\
$11$ & No & Yes & Yes & Alt. & --- & --- & Alt. & 6\\
$12$ & No & Yes & Yes & Alt./Lon. & --- & --- & Alt. & 7\\
$13$ & No & Yes & Yes & Alt./Lon./Lat. & --- & --- & Alt. & 8\\
$14$ & No & No & Yes & Alt. & Alt. & none & Alt. & 9\\
$15$ & No & No & Yes & Alt./Lon. & Alt./Lon. & none & Alt. & 11\\
$16$ & No & No & Yes & Alt./Lon./Lat. & Alt./Lon./Lat. & none & Alt. & 13\\
\end{tabular}
\end{table}

\end{document}